\documentclass[prl,showpacs,amsmath,twocolumn]{revtex4-1}
\usepackage[colorlinks=true,urlcolor=blue,citecolor=blue,linkcolor=blue]{hyperref}
\usepackage[sort&compress]{natbib}
\setcitestyle{numbers,square}

\usepackage[english]{babel}
\usepackage{setspace}
\usepackage{amssymb}
\usepackage{amsthm}
\usepackage{amscd}
\usepackage{rotating}
\usepackage{color}
\usepackage{graphics, graphicx}
\usepackage{psfrag}
\usepackage{mathcomp}
\usepackage{accents}

\begin{document}

\title{Thermometry for Laughlin States of Ultracold Atoms}
\author{P.T. Raum and V.W. Scarola}
\affiliation{Department of Physics, Virginia Tech, Blacksburg, Virginia 24061 USA}

\date{\today}

\begin{abstract}
Cooling atomic gases into strongly correlated quantum phases requires estimates of the entropy to perform thermometry and establish viability.  We construct an ansatz partition function for models of Laughlin states of atomic gases by combining high temperature series expansions with exact diagonalization. Using the ansatz we find that entropies required to observe Laughlin correlations with bosonic gases are within reach of current cooling capabilities.  \end{abstract}

\pacs{03.75.Hh,05.30.Jp,67.85.-d}

\maketitle

\noindent
 Observation of the superfluid-to-Mott transition \cite{jaksch:1998,greiner:2002} triggered interest in observing other strongly correlated states with ultracold atoms \cite{bloch:2008,lewenstein:2012}.  For example, a proposal \cite{hofstetter:2002} to probe models of high temperature superconductors with optical lattices led to efforts to emulate the controversial low temperature phase diagram of the Fermi-Hubbard model \cite{cho:2008,esslinger:2010}.  But cooling \cite{mckay:2011} proved to be a major obstacle.  Most atomic gas experiments are closed, to a good approximation.  The system entropy therefore determines the temperature.  Temperature is difficult to characterize in a strongly correlated regime because entropy-temperature relationships derive from non-trivial many-body effects.  Relatively recent theoretical work \cite{koetsier:2008,fuchs:2011,paiva:2011,kozik:2013} showed that the critical entropy to realize the best case scenario for emulation, the N\'{e}el state, lies below conventional evaporative cooling capabilities \cite{mckay:2011}.  In the case of the Fermi-Hubbard model, the entropy-temperature relationship proved to be unfavorable for realizing low temperatures at available entropies in conventional setups, although recent work with band engineering \cite{greif:2013} and trap shaping \cite{hart:2015} shows promise and progress with atomic gas microscopes \cite{parsons:2016,boll:2016,cheuk:2016,drewes:2016} has recently led to a report of long-range antiferromagnetism \cite{mazurenko:2016}.

A separate class of proposals seeks to realize fractional quantum Hall (FQH) states of atoms, particularly Laughlin states \cite{laughlin:1983b}, which, at low energies, have intriguing excitations with anyon statistics that define them as topological \cite{nayak:2008}.  These proposals rely on schemes to implement strong artificial magnetic fields (See Refs.~\cite{bloch:2008,cooper:2008,viefers:2008,salomon:2010,dalibard:2011,lewenstein:2012,goldman:2014a,goldman:2016} for reviews).  Experiments using rotation \cite{madison:2000,abo-shaeer:2001,bretin:2004,schweikhard:2004,tung:2006,gemelke:2010} or, more generally, methods to engineer the single particle phase \cite{lin:2009,lin:2009a,aidelsburger:2011,leblanc:2012,struck:2012,aidelsburger:2013,miyake:2013,atala:2014,stuhl:2015,kennedy:2015} have implemented strong magnetic fields.   But the thermodynamic relations needed for thermometry of large FQH systems are currently unknown.  

The absence of conventional Landau symmetry breaking precludes a Laughlin state critical temperature. But the energy gap establishes a Schottky-type peak separating an exponential from a power law temperature dependence in the heat capacity.  All estimates of equilibrium FQH observables, e.g., quantized edge current \cite{scarola:2007,zhao:2011,goldman:2013,atala:2014,stuhl:2015} or anyon statistics \cite{nayak:2008}, assume temperatures low enough to exponentially suppress excitations \cite{sarma:1997}.  We therefore use the peak as a necessary criterion for low temperature Laughlin correlations. 

 In a na\"{i}ve non-interacting model of gapped excitations \cite{supplmentaryinfo} an entropy per particle below $\approx0.29 k_{\text{B}}$ is needed to lower the temperature below the heat capacity peak.  This low entropy, if physically accurate, would imply that evaporative cooling methods cannot cool below the heat capacity peak since the experimental minimum is $\approx0.35 k_{\text{B}}$ for bosons \cite{mckay:2011}.  Entropies corresponding to temperatures below the peak are necessary.  A more accurate estimate of the entropy, incorporating the non-trivial excited state spectra of Laughlin states, is therefore needed to establish viability of low temperature FQH states with atomic gases.

Analyses of non-perturbative FQH models rely on a combination of numerics and ansatz theories.  Exact diagonalization has been used to compute the heat capacity over the entire temperature range but only for small systems \cite{yoshioka:1987,chakraborty:1997}.  Other work on the thermodynamics of infinite system sizes uses series expansions \cite{zheng:1994,tevosyan:1997,sawatdiaree:2000} but these studies are restricted to temperatures much larger than the gap.  Ansatz theories \cite{murthy:2003}, building on the success of the composite fermion (CF) wavefunctions \cite{jain:1989,jain:2007} at describing the low energy excitations \cite{jain:2005}, offer estimates for thermodynamic functions only at low temperatures.  A theory for the entropy-temperature relationship over the complete temperature range in large systems is necessary for guiding atomic gas experiments as they cool into FQH states.

We construct a theory of Laughlin state thermodynamics using an ansatz partition function designed to be exact at high temperatures, to capture the low temperature asymptotics, and to be straightforward to use, thus allowing characterization of the entire temperature range.  We establish a numerical validation procedure (which combines a high temperature series expansion with the stochastic trace method \cite{weisse:2006}) to compare our ansatz against exact results where possible.  Using our ansatz we incorporate FQH excitations \cite{jain:1989,jain:2005,jain:2007} to find that, in contrast to the na\"{i}ve estimate \cite{supplmentaryinfo}, entropies currently accessible with bosonic gases \cite{mckay:2011} are low enough to cool below the heat capacity peak.  Remarkably, the energy distribution of the excited states effectively lowers the temperature of the Schottky-type peak at fixed entropy in comparison to the na\"{i}ve estimate.  Our method can be used to construct thermodynamic functions of other FQH states.

\begin{figure}[t]
	\centering
	\hspace{-0.5cm}	
	\includegraphics[scale =.38]{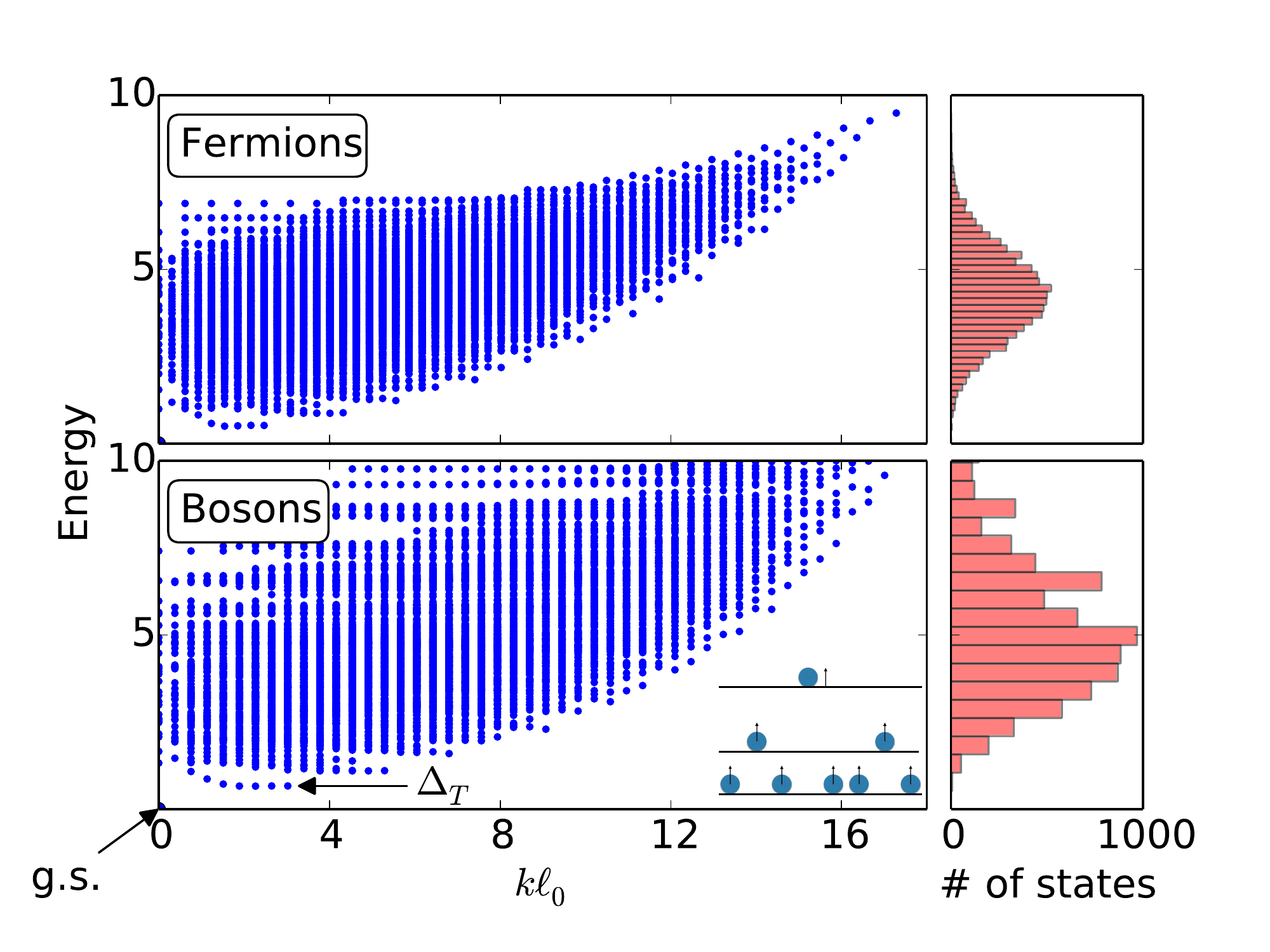}
	\vspace{-0.5cm}
	\caption {Energy of $N=8$ fermions (bosons) at $\nu=1/3$ $(\nu=1/2)$ as a function of total wavevector for $n=1$ ($n=0$) in Eq.~\eqref{eq_H} in the lowest Landau level.  The ground state is the $k=0$ Laughlin state set to zero energy.  $\Delta_T$ is the transport gap.  The schematic depicts example excitations.  The bosonic Laughlin state can be thought of as one filled level of CFs (bosons attached to one flux quantum).  Low energy excitations are CF particle-hole pairs modeled as excitations with energy $E_1\sim\Delta_T$.  We consider additional excitations at higher energies, e.g., $E_2$, where bosons are not necessarily bound to flux quanta.  The histograms show nearly Gaussian state counting.   }
	\label{fig_energy}
\end{figure}

\noindent
{\it Model:}  We consider a Hamiltonian of $N$ particles of mass $M$ in two-dimensions subjected to an artificial magnetic field oriented perpendicular to the $x-y$ plane:
\begin{eqnarray}
H&=&\sum_{i=1}^{N}\left[ \frac{\vert \mathbf{p}_{i}-q^*\mathbf{A}_{i}/c \vert^{2}}{2M}+\frac{M(\omega^2-\Omega^2)}{2}r_i^2\right] \nonumber \\ 
&+&b_{n}\sum_{i<j}^N\mathbf{\nabla}^{2n}\delta(\mathbf{r}_{i}-\mathbf{r}_{j}), 
\label{eq_H}
\end{eqnarray}
where $\mathbf{p}_i$ is the planar momentum, $\mathbf{A}_i= (\mathbf{B}^{*}\times\mathbf{r}_i)/2$ is the vector potential in the symmetric gauge, $\Omega\equiv q^*B^*/2M$ \cite{jain:2007,cooper:2008},  $\mathbf{r}_i=(x_i,y_i)$, and $B^*$ ($q^*$) is the artificial magnetic field (charge).  The effective magnetic length is $l_{0}=(\hbar c/q^*B^*)^{1/2}$.  In this gauge the concentric ring-like basis states define the disk geometry.  We assume a strong trap along the $z$-direction and an external parabolic confinement in the $x-y$ plane with a trapping frequency $\omega$.  To focus on bulk states we set $\omega=\Omega$ where the field cancels the effect of the trap \cite{jain:2007,cooper:2008} and discuss edge effects at the end.  We work in units of $l_0=k_B=1$.  The $s$-wave ($p$-wave) interaction for $n=0$ ($n=1$) generates repulsion and equates to the pseudopotential formulation \cite{haldane:1983}.  Setting $b_{n}=1$ defines our energy unit. 

The magnetic field must be large enough to restrict states to the lowest Landau level.  In this approximation the Laughlin state is the exact ground state \cite{haldane:1983,trugman:1985} of Eq.~\eqref{eq_H} for bosons (fermions) with $n=0$ $(n=1)$ at $\nu=1/2$ $(1/3)$, where $\nu$ is the filling factor, the number of particles per flux quanta.   In the following, when referring to bosons and fermions, we imply results at $\nu=1/2$ and $\nu=1/3$ with $n=0$ and $n=1$, respectively.

$H$ approximates several physical systems proposed for realizing FQH states with ultracold atoms.  We consider an atomic gas with a known entropy that is adiabatically loaded into a setup designed to generate $q^*B^*$.  For example, rotation generates $q^*B^*$ from the Coriolis effect \cite{mottelson:1999,cooper:1999,wilkin:2000,viefers:2000,regnault:2003,regnault:2004,cooper:2008}.  Artificial gauge fields in lattices offer another example. $H$ becomes accurate even in lattices when the flux through each unit cell is small (see, e.g., Ref~\cite{sorensen:2005}).

The Laughlin states form a subset of a larger class of states, the CF states, that accurately capture the low energy physics.  We think of a CF as a weakly interacting quasiparticle defined by attaching flux quanta to the original particles.  The Laughlin ground state becomes a filled effective level of CFs.  Low energy excitations are then particle-hole pairs of CFs which are $\mathcal{O}(1)$ different in energy from the ground state (See the schematic inset to Fig.~\ref{fig_energy}) and proliferate as temperatures increase to the heat capacity peak.  Near or above the peak, distinct excitations [$\mathcal{O}(N)$ different from the ground state] start to dominate.  
 
To study thermodynamics over the entire temperature range we use Eq.~\eqref{eq_H} to compute the energy in the spherical geometry, the geometry we use throughout.  The spherical geometry maps to the disk geometry in the $N\rightarrow\infty$ limit \cite{haldane:1983,fano:1986,wojs:1998} and allows us to focus on bulk states.  Fig.~\ref{fig_energy} shows a gap to a set of low energy modes, CF particle-hole pairs.  But the high energy states form a continuum which, as we will see, distinguishes the thermodynamics of Laughlin states from the na\"{i}ve model \cite{supplmentaryinfo}.

We take a statistical approach to incorporating the high energy continuum into the thermodynamics.  The histograms in Fig.~\ref{fig_energy} plot the distribution of energies.  We will rely on the observation that the continuum forms a nearly Gaussian distribution.  (Note that work in Ref.~\cite{wilkin:1998} implies that large vortices lead to the histogram peaks for bosons in Fig.~\ref{fig_energy}.)

Figure~\ref{fig_energy} shows only the excitations at fixed $N$.  In solids nearby particle reservoirs lead to addition or subtraction of additional particles (charged excitations) but in trapped atomic gases, particle number is essentially fixed. We therefore focus our analysis to fixed $N$ (neutral excitations).  We will also focus on uniform bulk states which implies that our results are relevant for systems with a small number of occupied edge states.  

$H$ separates into relative and center of mass coordinates allowing us to focus on excitations in the relative coordinates.  The total partition function in the canonical ensemble becomes:
$
Z_{\text{TOT}}=Z_{\text{CM}}\times Z,
$
where $Z_{\text{CM}}$ $(Z)$ is the partition function for the center of mass (relative) coordinates.  We approximate $Z$ for neutral excitations.

\noindent
{\it Method:} We construct an ansatz partition function that captures the exact thermodynamics at high temperatures and has the correct temperature dependence at low temperatures.  Considering the Gaussian-like energy distribution, the gapped spectrum, and the weakly interacting CF theory we arrive at: 
\begin{eqnarray}
Z_{A} \equiv \left(1+\sum_{m=1}^{3}g_{m}e^{-E_m/T} \right)^{N},
\label{eq_ansatz}
\end{eqnarray}
where $g_m$ and $E_m$ are fitting parameters.  The $m=1$ term in $Z_A$ defines the low $T$ dependence of a gapped spectrum.   The lowest energy excitation shown in Fig.~\ref{fig_energy} is $k$-dependent but we find that including $k$-dependence in $E_1$ does not alter our results.  We therefore ignore $k$-dependence 
and expect $E_1\sim\Delta_T$.  Our calculation of the gap is consistent with previous results obtained from the contact interaction \cite{nakajima:2003}.  (Different interactions alter the gap \cite{rezayi:2005,osterloh:2007,grusdt:2013}.)
Furthermore, $g_1$ approximates the degeneracy of the first excitation which should be unity within CF theory since there is one particle-hole excitation per $k$-state.   But we expect $g_1\gtrsim1$ since $g_1$ in our fitting renormalizes to account for additional nearby states in the many-body spectrum.  

\begin{table}
\center
\begin{tabular}{| l | l | l |}
	\hline
	& fermions & bosons \\ \hline
	$\Delta_T$ &  0.428(6) & 0.627(3) \\ \hline
	$\kappa_0/N$ & 1.899(1) & 1.899(1) \\ \hline
	$\kappa_1/N$ &  0.66647(2) & 1.005(4) \\ \hline
	$\kappa_2/N$ & 0.2974(1) & 2.28(2) \\ \hline
	$\kappa_3/N$ &  0.0856(1) & 25.2(3) \\ \hline
	$\kappa_4/N$ & 0.0016(1) & 6.8(2)$\times10^2$ \\ \hline
	$\kappa_5/N$ & 0.168(6) &3.3(1)$\times10^4$ \\ \hline
\end{tabular}
\caption{The first row shows the transport gap and the remaining rows the cumulants for fermions (bosons) at $\nu=1/3$ ($\nu=1/2$) for $n=1$ ($n=0$) in Eq.~\eqref{eq_H}.  All results are obtained by finite size scaling of exact numerical results to the thermodynamic limit \cite{supplmentaryinfo}. }
\label{table_cumulants}
\end{table}

We capture additional states in the high energy spectrum with $m>1$.   The inset to Fig.~\ref{fig_energy} schematically depicts another energy level, $E_{2}$.  At very high energies we expect states that do not involve flux attachment to play a role.   Small $g_m$ (and large $E_m$) implies diminishing impact of the $m^\text{th}$ level on the thermodynamics.

To fix parameters in the ansatz we consider an exact high temperature expansion related to the free energy:
\begin{eqnarray}
\log Z=\sum_{l=0}^{\infty}\kappa_{l}(-T)^{-l}/l!,
\end{eqnarray}
defined in terms of the cumulants, $\kappa_{l}$.  The lowest cumulants have simple interpretations, e.g., $\kappa_0$ is the log of the size of the Hilbert space.   Stirling's formula leads to $\kappa_0/N=3\log(3)-2\log(2)$ for $N\rightarrow\infty$.  Also, $\kappa_1/N=\text{Tr}(H)/N$ becomes $2\nu$ \cite{sawatdiaree:2000}.  Table~\ref{table_cumulants} shows the cumulants obtained from the stochastic trace method \cite{supplmentaryinfo}.

We use the cumulants from the high temperature expansion to fix the parameters in $Z_A$.  
We apply our method to bosons and reserve an analysis of fermions for future work because $n=1$ in Eq.~\eqref{eq_H} requires strong $p$-wave scattering which is experimentally challenging with alkali atoms.  
We use a  Gr\"{o}ebner basis fit to the high temperature limit.  We find: $g_1 = 5.53(3)$, $E_1 = 0.88(2)$, $g_2 = 0.22(3)$, $E_2 = 8.7(6)$, $g_3 = 0.0004(2)$, and $E_3 = 56.9(4.5)$.  The errors in our cumulant fits were propagated through our fitting procedure.  With these parameters we reproduce the exact high temperature expansion [up to $\mathcal{O}(T^{-6})$]. 

Reducing the number of fitting parameters tests accuracy.  By removing the $m=3$ terms in $Z_A$ we find that our results for fitting parameters, and therefore thermodynamic functions, change very little, e.g., the entropy changes by less than $1\%$.  
   
\noindent
{\it Thermodynamic Functions:} To test the validity of $Z_A$ we compare with exact diagonalization results for thermodynamic functions.  We compare directly on small system sizes and for the thermodynamic limit where possible. 

\begin{figure}[t]
\vspace{-0.75cm}
	\centering
	\includegraphics[scale =.45]{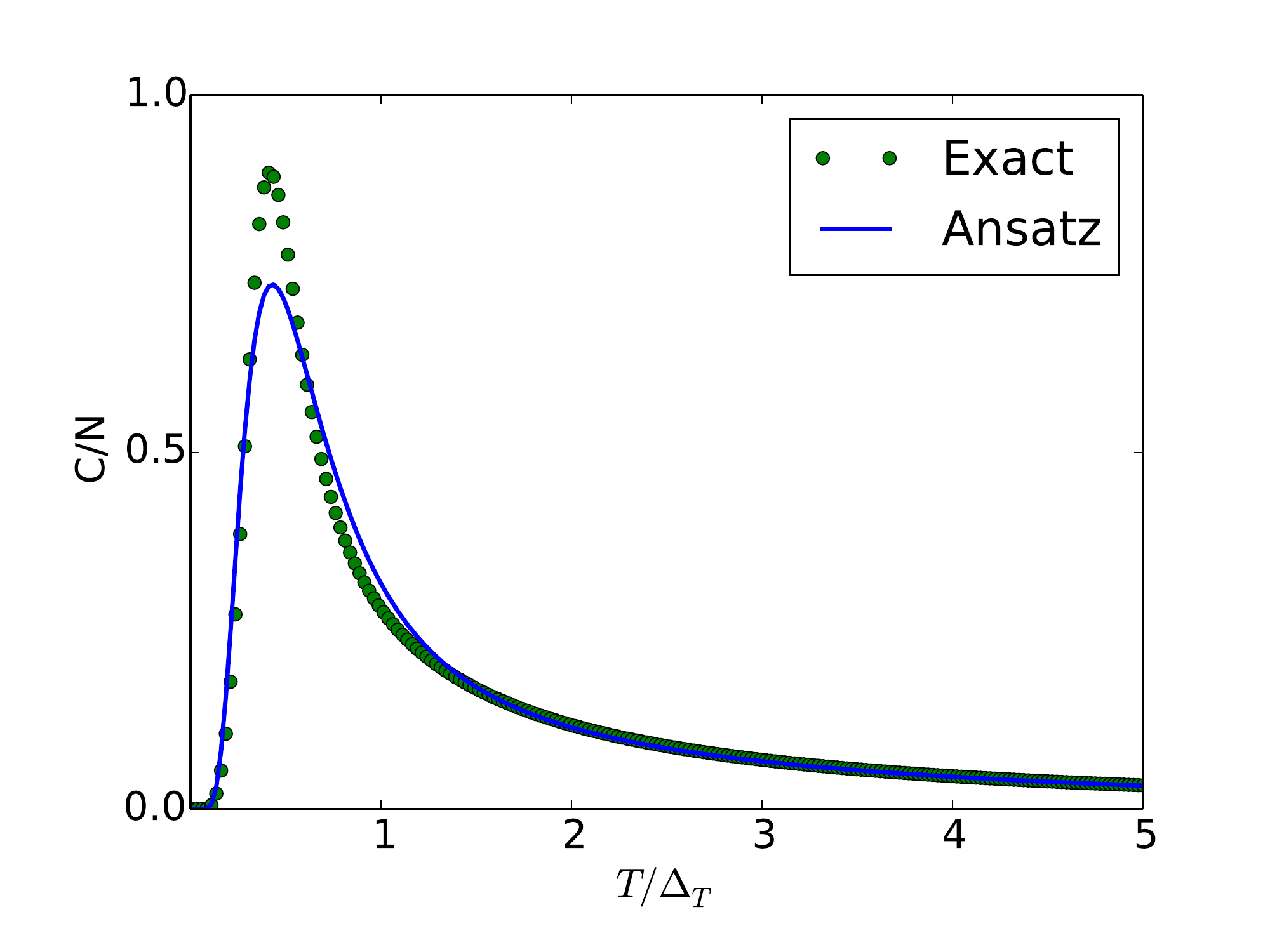}
	\vspace{-1cm}
	\caption {Heat capacity versus temperature for $N=8$ bosons.  The symbols are obtained from diagonalization of Eq.~\eqref{eq_H} and the solid line follows from the ansatz, Eq.~\eqref{eq_ansatz}.  The ansatz parameters were obtained by fitting the lowest six cumulants for $N=8$. The agreement at high $T$ follows by construction but the low $T$ agreement demonstrates the accuracy of the ansatz. }
	\label{fig_n8compare}
\end{figure}

Figure~\ref{fig_n8compare} plots the heat capacity, $C=T\partial S/\partial T$, where $S=\partial (T \log Z )/\partial T$ is the entropy.  We compare $C$ obtained from exact diagonalization and from the ansatz.  In this example comparison we see that both the high and low $T$ limits agree.  The low $T$ limit of the gapped system contains information about $g_1$, since for a system with a gap $\Delta_T$ and degeneracy $g$ we have:  $C/N\sim [ g(\Delta_T/ T)^{2}+\mathcal{O}(T^3)]e^{-\Delta_{T}/T}$.  We also find \cite{supplmentaryinfo} that increasing $N$ trends the peak from diagonalization toward the thermodynamic limit of the ansatz.

\begin{figure}[t]
\vspace{-0.75cm}
	\centering
	\includegraphics[scale =.45]{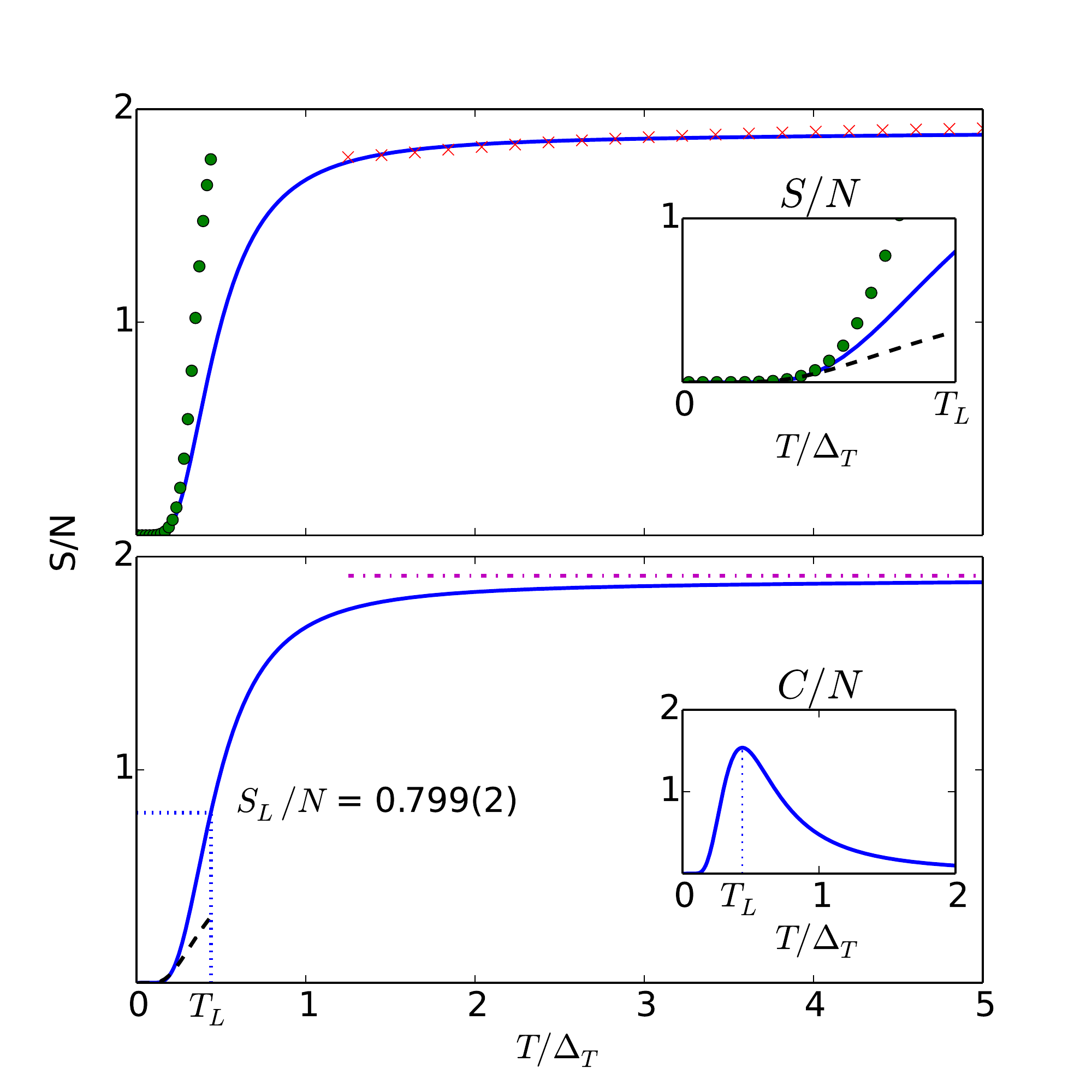}
	\vspace{-1cm}
	\caption {Top: Entropy per particle versus temperature for bosons.  The solid line is obtained from the ansatz, Eq.~\eqref{eq_ansatz}, with parameters derived from the cumulants in Table~\ref{table_cumulants}.  The circles (crosses) are obtained from finite $N$ extrapolations of the entropy at fixed low (high) $T$ \cite{supplmentaryinfo}.  The agreement between the symbols and the lines shows that the ansatz accurately captures the low and high $T$ limits.  Top Inset:  Low temperature zoom-in.  Bottom: The same but the dashed line plots the entropy per particle for the na\"{i}ve model \cite{supplmentaryinfo}.   The dot-dashed line plots the high $T$ limit from the first term of the cumulant expansion ($S=\kappa_0$).  Bottom Inset: Heat capacity from the ansatz in the thermodynamic limit.  The vertical dotted line denotes the temperature, $T_L$.  The horizontal dotted line in the main panel plots the corresponding entropy, $S_L$, determined from the ansatz.}
	\label{plot_SC}
\end{figure}

We also test $Z_A$ in the thermodynamic limit.  We study the entropy because of its importance in atomic gas thermometry.  We use exact diagonalization at finite $N$.  We fix $T$ and use finite size scaling to obtain the entropy at very low $T$ and very high $T$  \cite{supplmentaryinfo}.  For $T\sim\Delta_T$ we cannot predict an $N$ scaling function.  

The top panel of Fig.~\ref{plot_SC} plots a comparison between the entropy obtained from $Z_A$ and the exact results for finite-size scaling of the entropy.  Here we see that the ansatz agrees well with the extrapolations at low and high $T$.  But the symbols are not accurate for $T\sim\Delta_T$ because $N$-scaling breaks down here.  The bottom panel of Fig.~\ref{plot_SC} plots a comparison between the entropy obtained from $Z_A$ and two limits: the low $T$ limit obtained from the na\"{i}ve model \cite{supplmentaryinfo} and the high $T$ limit from the first term of the cumulant expansion ($S=\kappa_0$).  The deviation between the dashed and solid lines shows the importance of incorporating excitations beyond the na\"{i}ve model even for $T\sim \Delta_T/5$ \cite{supplmentaryinfo}.  In both panels we see that the ansatz extrapolates between both low and high $T$ limits thus allowing predictions for thermodynamic functions even for $T\sim\Delta_ T$.  

\noindent
{\it Laughlin Entropy:}  We use the validated ansatz to predict the entropy at which Laughlin correlations set in.  We consider a characteristic temperature, $T_L$, defined as the temperature at which the heat capacity peaks due to the energy gap.  The corresponding entropy is $S_L$.  

The inset of Fig.~\ref{plot_SC} plots the ansatz heat capacity versus temperature to reveal the location of the peak and therefore $T_L$.  The bottom panel of Fig.~\ref{plot_SC} shows that we find $S_L/N\approx0.799(2)$, where the error propagates from uncertainty in the ansatz fitting parameters.  We have  checked that artificial errors in the cumulants have linear impact on entropy, e.g., a $5\%$ variation of $\kappa_0$ led to a $5\%$ variation in $S_L$.  The $S_L$ we find is much larger than the na\"{i}ve estimate therefore showing the cooling effect of the continuum.  The entropy found here establishes a goal for experiments to reliably cool below $T_L$.  

\noindent
{\it Experimental Implications:}  We can compare our estimate for the entropy needed to cool into FQH states with bosons with current capabilities.  Evaporative cooling can reach entropies as low as $S/N\approx0.35$ \cite{mckay:2011} which, according to the ansatz, corresponds to $T\approx0.5 T_L$.  
We have therefore found that the entropy per particle required to lower the system temperature below $T_L$ is within reach.

Our estimates here only apply to neutral excitations within the bulk in the thermodynamic limit whereas edge effects can lower $T$ at fixed $S$ (adiabatic cooling) in finite sized systems.  
Introducing edges in a finite sized experiment should make the entropy budget more favorable \cite{greiner:2002}.  For $\Omega\neq \omega$, edge states interplay with parabolic trapping.  The (nearly) gapless edge states accommodate more entropy than the bulk [an $\mathcal{O}(\sqrt{N})$ impact on the entropy for a small number of edge modes].  Introducing edges should therefore adiabatically lower temperature. $S_L$ should then \emph{increase} once trap effects are included in modeling. 

Moving the filling away from $1/(1+p)$ introduces quasiparticles to allow additional cooling.  The topological nature of Laughlin states implies that the total entropy includes a factor due to quasiparticle degeneracy, yielding: $S_D+S$,  where $S_D=N_q\log(d)$, $d=\sqrt{p+1}$ is the quantum dimension, and $N_q$ is the number of additional quasiparticles causing deviation from filling $1/(1+p)$ \cite{nayak:2008}.  The $T$-independent $S_D$ term allows adiabatic cooling via quasiparticles \cite{gervais:2010}.  

\noindent
{\it Summary:}  We have constructed and validated a Laughlin state ansatz partition function for atomic gases.  Using our ansatz we find that the continuum of excited states alters the entropy-temperature relationship (in comparison to that of a na\"{i}ve gapped model) to reveal that currently attainable entropies with bosons are low enough to cool below the heat capacity peak.  Further work would allow thermometry on small system sizes by including edge states to compare theory with measurements, e.g., of pair correlation function \cite{gemelke:2010}.  Our results also have important implications for observing closely related topological states, e.g., chiral spin liquids and fractional Chern insulators, because of their direct connection to the bosonic Laughlin state \cite{kalmeyer:1987,yao:2013}. 

\begin{acknowledgments}
We acknowledge helpful comments from M. Peterson and B. De Marco and support from the AFOSR (FA9550-11-1-0313) and ARO (W911NF-16-1-0182).
\end{acknowledgments}

\bibliography{paper}

%merlin.mbs apsrev4-1.bst 2010-07-25 4.21a (PWD, AO, DPC) hacked
%Control: key (0)
%Control: author (8) initials jnrlst
%Control: editor formatted (1) identically to author
%Control: production of article title (-1) disabled
%Control: page (0) single
%Control: year (1) truncated
%Control: production of eprint (0) enabled
\begin{thebibliography}{77}%
\makeatletter
\providecommand \@ifxundefined [1]{%
 \@ifx{#1\undefined}
}%
\providecommand \@ifnum [1]{%
 \ifnum #1\expandafter \@firstoftwo
 \else \expandafter \@secondoftwo
 \fi
}%
\providecommand \@ifx [1]{%
 \ifx #1\expandafter \@firstoftwo
 \else \expandafter \@secondoftwo
 \fi
}%
\providecommand \natexlab [1]{#1}%
\providecommand \enquote  [1]{``#1''}%
\providecommand \bibnamefont  [1]{#1}%
\providecommand \bibfnamefont [1]{#1}%
\providecommand \citenamefont [1]{#1}%
\providecommand \href@noop [0]{\@secondoftwo}%
\providecommand \href [0]{\begingroup \@sanitize@url \@href}%
\providecommand \@href[1]{\@@startlink{#1}\@@href}%
\providecommand \@@href[1]{\endgroup#1\@@endlink}%
\providecommand \@sanitize@url [0]{\catcode `\\12\catcode `\$12\catcode
  `\&12\catcode `\#12\catcode `\^12\catcode `\_12\catcode `\%12\relax}%
\providecommand \@@startlink[1]{}%
\providecommand \@@endlink[0]{}%
\providecommand \url  [0]{\begingroup\@sanitize@url \@url }%
\providecommand \@url [1]{\endgroup\@href {#1}{\urlprefix }}%
\providecommand \urlprefix  [0]{URL }%
\providecommand \Eprint [0]{\href }%
\providecommand \doibase [0]{http://dx.doi.org/}%
\providecommand \selectlanguage [0]{\@gobble}%
\providecommand \bibinfo  [0]{\@secondoftwo}%
\providecommand \bibfield  [0]{\@secondoftwo}%
\providecommand \translation [1]{[#1]}%
\providecommand \BibitemOpen [0]{}%
\providecommand \bibitemStop [0]{}%
\providecommand \bibitemNoStop [0]{.\EOS\space}%
\providecommand \EOS [0]{\spacefactor3000\relax}%
\providecommand \BibitemShut  [1]{\csname bibitem#1\endcsname}%
\let\auto@bib@innerbib\@empty
%</preamble>
\bibitem [{\citenamefont {Jaksch}\ \emph {et~al.}(1998)\citenamefont {Jaksch},
  \citenamefont {Bruder}, \citenamefont {Cirac}, \citenamefont {Gardiner},\
  and\ \citenamefont {Zoller}}]{jaksch:1998}%
  \BibitemOpen
  \bibfield  {author} {\bibinfo {author} {\bibfnamefont {D.}~\bibnamefont
  {Jaksch}}, \bibinfo {author} {\bibfnamefont {C.}~\bibnamefont {Bruder}},
  \bibinfo {author} {\bibfnamefont {J.~I.}\ \bibnamefont {Cirac}}, \bibinfo
  {author} {\bibfnamefont {C.~W.}\ \bibnamefont {Gardiner}}, \ and\ \bibinfo
  {author} {\bibfnamefont {P.}~\bibnamefont {Zoller}},\ }\href
  {http://dx.doi.org/10.1103/PhysRevLett.81.3108} {\bibfield  {journal}
  {\bibinfo  {journal} {Phys. Rev. Lett.}\ }\textbf {\bibinfo {volume} {81}},\
  \bibinfo {pages} {3108} (\bibinfo {year} {1998})}\BibitemShut {NoStop}%
\bibitem [{\citenamefont {Greiner}\ \emph {et~al.}(2002)\citenamefont
  {Greiner}, \citenamefont {Mandel}, \citenamefont {Esslinger}, \citenamefont
  {Hansch},\ and\ \citenamefont {Bloch}}]{greiner:2002}%
  \BibitemOpen
  \bibfield  {author} {\bibinfo {author} {\bibfnamefont {M.}~\bibnamefont
  {Greiner}}, \bibinfo {author} {\bibfnamefont {O.}~\bibnamefont {Mandel}},
  \bibinfo {author} {\bibfnamefont {T.}~\bibnamefont {Esslinger}}, \bibinfo
  {author} {\bibfnamefont {T.~W.}\ \bibnamefont {Hansch}}, \ and\ \bibinfo
  {author} {\bibfnamefont {I.}~\bibnamefont {Bloch}},\ }\href
  {http://dx.doi.org/10.1038/415039a} {\bibfield  {journal} {\bibinfo
  {journal} {Nature}\ }\textbf {\bibinfo {volume} {415}},\ \bibinfo {pages}
  {39} (\bibinfo {year} {2002})}\BibitemShut {NoStop}%
\bibitem [{\citenamefont {Bloch}\ \emph {et~al.}(2008)\citenamefont {Bloch},
  \citenamefont {Dalibard},\ and\ \citenamefont {Zwerger}}]{bloch:2008}%
  \BibitemOpen
  \bibfield  {author} {\bibinfo {author} {\bibfnamefont {I.}~\bibnamefont
  {Bloch}}, \bibinfo {author} {\bibfnamefont {J.}~\bibnamefont {Dalibard}}, \
  and\ \bibinfo {author} {\bibfnamefont {W.}~\bibnamefont {Zwerger}},\ }\href
  {http://dx.doi.org/10.1103/RevModPhys.80.885} {\bibfield  {journal} {\bibinfo
   {journal} {Rev. Mod. Phys.}\ }\textbf {\bibinfo {volume} {80}},\ \bibinfo
  {pages} {885} (\bibinfo {year} {2008})}\BibitemShut {NoStop}%
\bibitem [{\citenamefont {Lewenstein}\ \emph {et~al.}(2012)\citenamefont
  {Lewenstein}, \citenamefont {Sanpera},\ and\ \citenamefont
  {Ahufinger}}]{lewenstein:2012}%
  \BibitemOpen
  \bibfield  {author} {\bibinfo {author} {\bibfnamefont {M.}~\bibnamefont
  {Lewenstein}}, \bibinfo {author} {\bibfnamefont {A.}~\bibnamefont {Sanpera}},
  \ and\ \bibinfo {author} {\bibfnamefont {V.}~\bibnamefont {Ahufinger}},\
  }\href@noop {} {\emph {\bibinfo {title} {Ultracold Atoms in Optical
  Lattices:Simulating quantum many-body systems}}}\ (\bibinfo  {publisher} {OUP
  Oxford},\ \bibinfo {year} {2012})\BibitemShut {NoStop}%
\bibitem [{\citenamefont {Hofstetter}\ \emph {et~al.}(2002)\citenamefont
  {Hofstetter}, \citenamefont {Cirac}, \citenamefont {Zoller}, \citenamefont
  {Demler},\ and\ \citenamefont {Lukin}}]{hofstetter:2002}%
  \BibitemOpen
  \bibfield  {author} {\bibinfo {author} {\bibfnamefont {W.}~\bibnamefont
  {Hofstetter}}, \bibinfo {author} {\bibfnamefont {J.~I.}\ \bibnamefont
  {Cirac}}, \bibinfo {author} {\bibfnamefont {P.}~\bibnamefont {Zoller}},
  \bibinfo {author} {\bibfnamefont {E.}~\bibnamefont {Demler}}, \ and\ \bibinfo
  {author} {\bibfnamefont {M.~D.}\ \bibnamefont {Lukin}},\ }\href {\doibase
  10.1103/PhysRevLett.89.220407} {\bibfield  {journal} {\bibinfo  {journal}
  {Phys. Rev. Lett.}\ }\textbf {\bibinfo {volume} {89}},\ \bibinfo {pages}
  {220407} (\bibinfo {year} {2002})}\BibitemShut {NoStop}%
\bibitem [{\citenamefont {Cho}(2008)}]{cho:2008}%
  \BibitemOpen
  \bibfield  {author} {\bibinfo {author} {\bibfnamefont {A.}~\bibnamefont
  {Cho}},\ }\href {http://dx.doi.org/10.1126/science.320.5874.312} {\bibfield
  {journal} {\bibinfo  {journal} {Science}\ }\textbf {\bibinfo {volume}
  {320}},\ \bibinfo {pages} {312} (\bibinfo {year} {2008})}\BibitemShut
  {NoStop}%
\bibitem [{\citenamefont {Esslinger}(2010)}]{esslinger:2010}%
  \BibitemOpen
  \bibfield  {author} {\bibinfo {author} {\bibfnamefont {T.}~\bibnamefont
  {Esslinger}},\ }\href {\doibase 10.1146/annurev-conmatphys-070909-104059}
  {\bibfield  {journal} {\bibinfo  {journal} {Ann. Rev. of Condens. Matt.
  Phys.}\ }\textbf {\bibinfo {volume} {1}},\ \bibinfo {pages} {129} (\bibinfo
  {year} {2010})}\BibitemShut {NoStop}%
\bibitem [{\citenamefont {{McKay}}\ and\ \citenamefont
  {{DeMarco}}(2011)}]{mckay:2011}%
  \BibitemOpen
  \bibfield  {author} {\bibinfo {author} {\bibfnamefont {D.~C.}\ \bibnamefont
  {{McKay}}}\ and\ \bibinfo {author} {\bibfnamefont {B.}~\bibnamefont
  {{DeMarco}}},\ }\href {\doibase 10.1088/0034-4885/74/5/054401} {\bibfield
  {journal} {\bibinfo  {journal} {Rep. Prog. Phys.}\ }\textbf {\bibinfo
  {volume} {74}},\ \bibinfo {eid} {054401} (\bibinfo {year}
  {2011})}\BibitemShut {NoStop}%
\bibitem [{\citenamefont {Koetsier}\ \emph {et~al.}(2008)\citenamefont
  {Koetsier}, \citenamefont {Duine}, \citenamefont {Bloch},\ and\ \citenamefont
  {Stoof}}]{koetsier:2008}%
  \BibitemOpen
  \bibfield  {author} {\bibinfo {author} {\bibfnamefont {A.}~\bibnamefont
  {Koetsier}}, \bibinfo {author} {\bibfnamefont {R.~A.}\ \bibnamefont {Duine}},
  \bibinfo {author} {\bibfnamefont {I.}~\bibnamefont {Bloch}}, \ and\ \bibinfo
  {author} {\bibfnamefont {H.~T.~C.}\ \bibnamefont {Stoof}},\ }\href {\doibase
  10.1103/PhysRevA.77.023623} {\bibfield  {journal} {\bibinfo  {journal} {Phys.
  Rev. A}\ }\textbf {\bibinfo {volume} {77}},\ \bibinfo {pages} {023623}
  (\bibinfo {year} {2008})}\BibitemShut {NoStop}%
\bibitem [{\citenamefont {Fuchs}\ \emph {et~al.}(2011)\citenamefont {Fuchs},
  \citenamefont {Gull}, \citenamefont {Pollet}, \citenamefont {Burovski},
  \citenamefont {Kozik}, \citenamefont {Pruschke},\ and\ \citenamefont
  {Troyer}}]{fuchs:2011}%
  \BibitemOpen
  \bibfield  {author} {\bibinfo {author} {\bibfnamefont {S.}~\bibnamefont
  {Fuchs}}, \bibinfo {author} {\bibfnamefont {E.}~\bibnamefont {Gull}},
  \bibinfo {author} {\bibfnamefont {L.}~\bibnamefont {Pollet}}, \bibinfo
  {author} {\bibfnamefont {E.}~\bibnamefont {Burovski}}, \bibinfo {author}
  {\bibfnamefont {E.}~\bibnamefont {Kozik}}, \bibinfo {author} {\bibfnamefont
  {T.}~\bibnamefont {Pruschke}}, \ and\ \bibinfo {author} {\bibfnamefont
  {M.}~\bibnamefont {Troyer}},\ }\href {\doibase
  10.1103/physrevlett.106.030401} {\bibfield  {journal} {\bibinfo  {journal}
  {Phys. Rev. Lett.}\ }\textbf {\bibinfo {volume} {106}},\ \bibinfo {pages}
  {030401} (\bibinfo {year} {2011})}\BibitemShut {NoStop}%
\bibitem [{\citenamefont {{Paiva}}\ \emph {et~al.}(2011)\citenamefont
  {{Paiva}}, \citenamefont {{Loh}}, \citenamefont {{Randeria}}, \citenamefont
  {{Scalettar}},\ and\ \citenamefont {{Trivedi}}}]{paiva:2011}%
  \BibitemOpen
  \bibfield  {author} {\bibinfo {author} {\bibfnamefont {T.}~\bibnamefont
  {{Paiva}}}, \bibinfo {author} {\bibfnamefont {Y.~L.}\ \bibnamefont {{Loh}}},
  \bibinfo {author} {\bibfnamefont {M.}~\bibnamefont {{Randeria}}}, \bibinfo
  {author} {\bibfnamefont {R.~T.}\ \bibnamefont {{Scalettar}}}, \ and\ \bibinfo
  {author} {\bibfnamefont {N.}~\bibnamefont {{Trivedi}}},\ }\href {\doibase
  10.1103/PhysRevLett.107.086401} {\bibfield  {journal} {\bibinfo  {journal}
  {Phys. Rev. Lett.}\ }\textbf {\bibinfo {volume} {107}},\ \bibinfo {eid}
  {086401} (\bibinfo {year} {2011})}\BibitemShut {NoStop}%
\bibitem [{\citenamefont {Kozik}\ \emph {et~al.}(2013)\citenamefont {Kozik},
  \citenamefont {Burovski}, \citenamefont {Scarola},\ and\ \citenamefont
  {Troyer}}]{kozik:2013}%
  \BibitemOpen
  \bibfield  {author} {\bibinfo {author} {\bibfnamefont {E.}~\bibnamefont
  {Kozik}}, \bibinfo {author} {\bibfnamefont {E.}~\bibnamefont {Burovski}},
  \bibinfo {author} {\bibfnamefont {V.~W.}\ \bibnamefont {Scarola}}, \ and\
  \bibinfo {author} {\bibfnamefont {M.}~\bibnamefont {Troyer}},\ }\href
  {\doibase 10.1103/PhysRevB.87.205102} {\bibfield  {journal} {\bibinfo
  {journal} {Phys. Rev. B}\ }\textbf {\bibinfo {volume} {87}},\ \bibinfo
  {pages} {205102} (\bibinfo {year} {2013})}\BibitemShut {NoStop}%
\bibitem [{\citenamefont {Greif}\ \emph {et~al.}(2013)\citenamefont {Greif},
  \citenamefont {Uehlinger}, \citenamefont {Jotzu}, \citenamefont {Tarruell},\
  and\ \citenamefont {Esslinger}}]{greif:2013}%
  \BibitemOpen
  \bibfield  {author} {\bibinfo {author} {\bibfnamefont {D.}~\bibnamefont
  {Greif}}, \bibinfo {author} {\bibfnamefont {T.}~\bibnamefont {Uehlinger}},
  \bibinfo {author} {\bibfnamefont {G.}~\bibnamefont {Jotzu}}, \bibinfo
  {author} {\bibfnamefont {L.}~\bibnamefont {Tarruell}}, \ and\ \bibinfo
  {author} {\bibfnamefont {T.}~\bibnamefont {Esslinger}},\ }\href {\doibase
  10.1126/science.1236362} {\bibfield  {journal} {\bibinfo  {journal}
  {Science}\ }\textbf {\bibinfo {volume} {340}},\ \bibinfo {pages} {1307}
  (\bibinfo {year} {2013})}\BibitemShut {NoStop}%
\bibitem [{\citenamefont {Hart}\ \emph {et~al.}(2015)\citenamefont {Hart},
  \citenamefont {Duarte}, \citenamefont {Yang}, \citenamefont {Liu},
  \citenamefont {Paiva}, \citenamefont {Khatami}, \citenamefont {Scalettar},
  \citenamefont {Trivedi}, \citenamefont {Huse},\ and\ \citenamefont
  {Hulet}}]{hart:2015}%
  \BibitemOpen
  \bibfield  {author} {\bibinfo {author} {\bibfnamefont {R.~A.}\ \bibnamefont
  {Hart}}, \bibinfo {author} {\bibfnamefont {P.~M.}\ \bibnamefont {Duarte}},
  \bibinfo {author} {\bibfnamefont {T.}~\bibnamefont {Yang}}, \bibinfo {author}
  {\bibfnamefont {X.}~\bibnamefont {Liu}}, \bibinfo {author} {\bibfnamefont
  {T.}~\bibnamefont {Paiva}}, \bibinfo {author} {\bibfnamefont
  {E.}~\bibnamefont {Khatami}}, \bibinfo {author} {\bibfnamefont {R.~T.}\
  \bibnamefont {Scalettar}}, \bibinfo {author} {\bibfnamefont {N.}~\bibnamefont
  {Trivedi}}, \bibinfo {author} {\bibfnamefont {D.~A.}\ \bibnamefont {Huse}}, \
  and\ \bibinfo {author} {\bibfnamefont {R.~G.}\ \bibnamefont {Hulet}},\ }\href
  {\doibase 10.1038/nature14223} {\bibfield  {journal} {\bibinfo  {journal}
  {Nature}\ }\textbf {\bibinfo {volume} {519}},\ \bibinfo {pages} {211}
  (\bibinfo {year} {2015})}\BibitemShut {NoStop}%
\bibitem [{\citenamefont {Parsons}\ \emph {et~al.}(2016)\citenamefont
  {Parsons}, \citenamefont {Mazurenko}, \citenamefont {Chiu}, \citenamefont
  {Ji}, \citenamefont {Greif},\ and\ \citenamefont {Greiner}}]{parsons:2016}%
  \BibitemOpen
  \bibfield  {author} {\bibinfo {author} {\bibfnamefont {M.~F.}\ \bibnamefont
  {Parsons}}, \bibinfo {author} {\bibfnamefont {A.}~\bibnamefont {Mazurenko}},
  \bibinfo {author} {\bibfnamefont {C.~S.}\ \bibnamefont {Chiu}}, \bibinfo
  {author} {\bibfnamefont {G.}~\bibnamefont {Ji}}, \bibinfo {author}
  {\bibfnamefont {D.}~\bibnamefont {Greif}}, \ and\ \bibinfo {author}
  {\bibfnamefont {M.}~\bibnamefont {Greiner}},\ }\href {\doibase
  10.1126/science.aag1430} {\bibfield  {journal} {\bibinfo  {journal}
  {Science}\ }\textbf {\bibinfo {volume} {353}},\ \bibinfo {pages} {1253}
  (\bibinfo {year} {2016})}\BibitemShut {NoStop}%
\bibitem [{\citenamefont {Boll}\ \emph {et~al.}(2016)\citenamefont {Boll},
  \citenamefont {Hilker}, \citenamefont {Salomon}, \citenamefont {Omran},
  \citenamefont {Nespolo}, \citenamefont {Pollet}, \citenamefont {Bloch},\ and\
  \citenamefont {Gross}}]{boll:2016}%
  \BibitemOpen
  \bibfield  {author} {\bibinfo {author} {\bibfnamefont {M.}~\bibnamefont
  {Boll}}, \bibinfo {author} {\bibfnamefont {T.~A.}\ \bibnamefont {Hilker}},
  \bibinfo {author} {\bibfnamefont {G.}~\bibnamefont {Salomon}}, \bibinfo
  {author} {\bibfnamefont {A.}~\bibnamefont {Omran}}, \bibinfo {author}
  {\bibfnamefont {J.}~\bibnamefont {Nespolo}}, \bibinfo {author} {\bibfnamefont
  {L.}~\bibnamefont {Pollet}}, \bibinfo {author} {\bibfnamefont
  {I.}~\bibnamefont {Bloch}}, \ and\ \bibinfo {author} {\bibfnamefont
  {C.}~\bibnamefont {Gross}},\ }\href {\doibase 10.1126/science.aag1635}
  {\bibfield  {journal} {\bibinfo  {journal} {Science}\ }\textbf {\bibinfo
  {volume} {353}},\ \bibinfo {pages} {1257} (\bibinfo {year}
  {2016})}\BibitemShut {NoStop}%
\bibitem [{\citenamefont {Cheuk}\ \emph {et~al.}(2016)\citenamefont {Cheuk},
  \citenamefont {Nichols}, \citenamefont {Lawrence}, \citenamefont {Okan},
  \citenamefont {Zhang}, \citenamefont {Khatami}, \citenamefont {Trivedi},
  \citenamefont {Paiva}, \citenamefont {Rigol},\ and\ \citenamefont
  {Zwierlein}}]{cheuk:2016}%
  \BibitemOpen
  \bibfield  {author} {\bibinfo {author} {\bibfnamefont {L.~W.}\ \bibnamefont
  {Cheuk}}, \bibinfo {author} {\bibfnamefont {M.~A.}\ \bibnamefont {Nichols}},
  \bibinfo {author} {\bibfnamefont {K.~R.}\ \bibnamefont {Lawrence}}, \bibinfo
  {author} {\bibfnamefont {M.}~\bibnamefont {Okan}}, \bibinfo {author}
  {\bibfnamefont {H.}~\bibnamefont {Zhang}}, \bibinfo {author} {\bibfnamefont
  {E.}~\bibnamefont {Khatami}}, \bibinfo {author} {\bibfnamefont
  {N.}~\bibnamefont {Trivedi}}, \bibinfo {author} {\bibfnamefont
  {T.}~\bibnamefont {Paiva}}, \bibinfo {author} {\bibfnamefont
  {M.}~\bibnamefont {Rigol}}, \ and\ \bibinfo {author} {\bibfnamefont {M.~W.}\
  \bibnamefont {Zwierlein}},\ }\href {\doibase 10.1126/science.aag3349}
  {\bibfield  {journal} {\bibinfo  {journal} {Science}\ }\textbf {\bibinfo
  {volume} {353}},\ \bibinfo {pages} {1260} (\bibinfo {year}
  {2016})}\BibitemShut {NoStop}%
\bibitem [{\citenamefont {Drewes}\ \emph {et~al.}(2016)\citenamefont {Drewes},
  \citenamefont {Miller}, \citenamefont {Cocchi}, \citenamefont {Chan},
  \citenamefont {Pertot}, \citenamefont {Brennecke},\ and\ \citenamefont
  {K{\"o}hl}}]{drewes:2016}%
  \BibitemOpen
  \bibfield  {author} {\bibinfo {author} {\bibfnamefont {J.}~\bibnamefont
  {Drewes}}, \bibinfo {author} {\bibfnamefont {L.}~\bibnamefont {Miller}},
  \bibinfo {author} {\bibfnamefont {E.}~\bibnamefont {Cocchi}}, \bibinfo
  {author} {\bibfnamefont {C.}~\bibnamefont {Chan}}, \bibinfo {author}
  {\bibfnamefont {D.}~\bibnamefont {Pertot}}, \bibinfo {author} {\bibfnamefont
  {F.}~\bibnamefont {Brennecke}}, \ and\ \bibinfo {author} {\bibfnamefont
  {M.}~\bibnamefont {K{\"o}hl}},\ }\href@noop {} {\bibfield  {journal}
  {\bibinfo  {journal} {arXiv preprint arXiv:1607.00392}\ } (\bibinfo {year}
  {2016})}\BibitemShut {NoStop}%
\bibitem [{\citenamefont {{{Mazurenko}, C.~S. {Chiu}, G. {Ji}, M.~F. {Parsons},
  M. {Kan{\'a}sz-Nagy}, R. {Schmidt}, F. {Grusdt}, E. {Demler}, D. {Greif}, and
  M. {Greiner}}}(2016)}]{mazurenko:2016}%
  \BibitemOpen
  \bibfield  {author} {\bibinfo {author} {\bibfnamefont {A.}~\bibnamefont
  {{{Mazurenko}, C.~S. {Chiu}, G. {Ji}, M.~F. {Parsons}, M. {Kan{\'a}sz-Nagy},
  R. {Schmidt}, F. {Grusdt}, E. {Demler}, D. {Greif}, and M. {Greiner}}}},\
  }\href {https://arxiv.org/abs/1612.08436} {\bibfield  {journal} {\bibinfo
  {journal} {arXiv:1607.00392}\ } (\bibinfo {year} {2016})}\BibitemShut
  {NoStop}%
\bibitem [{\citenamefont {Laughlin}(1983)}]{laughlin:1983b}%
  \BibitemOpen
  \bibfield  {author} {\bibinfo {author} {\bibfnamefont {R.}~\bibnamefont
  {Laughlin}},\ }\href {http://dx.doi.org/10.1103/PhysRevLett.50.1395}
  {\bibfield  {journal} {\bibinfo  {journal} {Phys. Rev. Lett.}\ }\textbf
  {\bibinfo {volume} {50}},\ \bibinfo {pages} {1395} (\bibinfo {year}
  {1983})}\BibitemShut {NoStop}%
\bibitem [{\citenamefont {Nayak}\ \emph {et~al.}(2008)\citenamefont {Nayak},
  \citenamefont {Simon}, \citenamefont {Stern}, \citenamefont {Freedman},\ and\
  \citenamefont {Das~Sarma}}]{nayak:2008}%
  \BibitemOpen
  \bibfield  {author} {\bibinfo {author} {\bibfnamefont {C.}~\bibnamefont
  {Nayak}}, \bibinfo {author} {\bibfnamefont {S.~H.}\ \bibnamefont {Simon}},
  \bibinfo {author} {\bibfnamefont {A.}~\bibnamefont {Stern}}, \bibinfo
  {author} {\bibfnamefont {M.}~\bibnamefont {Freedman}}, \ and\ \bibinfo
  {author} {\bibfnamefont {S.}~\bibnamefont {Das~Sarma}},\ }\href {\doibase
  10.1103/RevModPhys.80.1083} {\bibfield  {journal} {\bibinfo  {journal} {Rev.
  Mod. Phys.}\ }\textbf {\bibinfo {volume} {80}},\ \bibinfo {pages} {1083}
  (\bibinfo {year} {2008})}\BibitemShut {NoStop}%
\bibitem [{\citenamefont {Cooper}(2008)}]{cooper:2008}%
  \BibitemOpen
  \bibfield  {author} {\bibinfo {author} {\bibfnamefont {N.~R.}\ \bibnamefont
  {Cooper}},\ }\href {\doibase 10.1080/00018730802564122} {\bibfield  {journal}
  {\bibinfo  {journal} {Adv. Phys.}\ }\textbf {\bibinfo {volume} {57}},\
  \bibinfo {pages} {539} (\bibinfo {year} {2008})}\BibitemShut {NoStop}%
\bibitem [{\citenamefont {Viefers}(2008)}]{viefers:2008}%
  \BibitemOpen
  \bibfield  {author} {\bibinfo {author} {\bibfnamefont {S.}~\bibnamefont
  {Viefers}},\ }\href {\doibase 10.1088/0953-8984/20/12/123202} {\bibfield
  {journal} {\bibinfo  {journal} {J. Phys.: Condens. Matter}\ }\textbf
  {\bibinfo {volume} {20}},\ \bibinfo {pages} {123202} (\bibinfo {year}
  {2008})}\BibitemShut {NoStop}%
\bibitem [{\citenamefont {Salomon}\ \emph {et~al.}(2013)\citenamefont
  {Salomon}, \citenamefont {Shlyapnikov},\ and\ \citenamefont
  {Cugliandolo}}]{salomon:2010}%
  \BibitemOpen
  \bibfield  {author} {\bibinfo {author} {\bibfnamefont {C.}~\bibnamefont
  {Salomon}}, \bibinfo {author} {\bibfnamefont {G.~V.}\ \bibnamefont
  {Shlyapnikov}}, \ and\ \bibinfo {author} {\bibfnamefont {L.~F.}\ \bibnamefont
  {Cugliandolo}},\ }\href@noop {} {\emph {\bibinfo {title} {Many-Body Physics
  with Ultracold Gases: Lecture Notes of the Les Houches Summer School: Volume
  94, July 2010}}}\ (\bibinfo  {publisher} {Oxford University Press},\ \bibinfo
  {year} {2013})\BibitemShut {NoStop}%
\bibitem [{\citenamefont {Dalibard}\ \emph {et~al.}(2011)\citenamefont
  {Dalibard}, \citenamefont {Gerbier}, \citenamefont
  {Juzeli\ifmmode~\bar{u}\else \={u}\fi{}nas},\ and\ \citenamefont
  {{\"O}hberg}}]{dalibard:2011}%
  \BibitemOpen
  \bibfield  {author} {\bibinfo {author} {\bibfnamefont {J.}~\bibnamefont
  {Dalibard}}, \bibinfo {author} {\bibfnamefont {F.}~\bibnamefont {Gerbier}},
  \bibinfo {author} {\bibfnamefont {G.}~\bibnamefont
  {Juzeli\ifmmode~\bar{u}\else \={u}\fi{}nas}}, \ and\ \bibinfo {author}
  {\bibfnamefont {P.}~\bibnamefont {{\"O}hberg}},\ }\href {\doibase
  10.1103/RevModPhys.83.1523} {\bibfield  {journal} {\bibinfo  {journal} {Rev.
  Mod. Phys.}\ }\textbf {\bibinfo {volume} {83}},\ \bibinfo {pages} {1523}
  (\bibinfo {year} {2011})}\BibitemShut {NoStop}%
\bibitem [{\citenamefont {Goldman}\ \emph {et~al.}(2014)\citenamefont
  {Goldman}, \citenamefont {Juzelinas}, \citenamefont {Ohberg},\ and\
  \citenamefont {Spielman}}]{goldman:2014a}%
  \BibitemOpen
  \bibfield  {author} {\bibinfo {author} {\bibfnamefont {N.}~\bibnamefont
  {Goldman}}, \bibinfo {author} {\bibfnamefont {G.}~\bibnamefont {Juzelinas}},
  \bibinfo {author} {\bibfnamefont {P.}~\bibnamefont {Ohberg}}, \ and\ \bibinfo
  {author} {\bibfnamefont {I.~B.}\ \bibnamefont {Spielman}},\ }\href
  {http://stacks.iop.org/0034-4885/77/i=12/a=126401} {\bibfield  {journal}
  {\bibinfo  {journal} {Rep. Prog. Phys.}\ }\textbf {\bibinfo {volume} {77}},\
  \bibinfo {pages} {126401} (\bibinfo {year} {2014})}\BibitemShut {NoStop}%
\bibitem [{\citenamefont {Goldman}\ \emph {et~al.}(2016)\citenamefont
  {Goldman}, \citenamefont {Budich},\ and\ \citenamefont
  {Zoller}}]{goldman:2016}%
  \BibitemOpen
  \bibfield  {author} {\bibinfo {author} {\bibfnamefont {N.}~\bibnamefont
  {Goldman}}, \bibinfo {author} {\bibfnamefont {J.~C.}\ \bibnamefont {Budich}},
  \ and\ \bibinfo {author} {\bibfnamefont {P.}~\bibnamefont {Zoller}},\ }\href
  {\doibase 10.1038/nphys3803} {\bibfield  {journal} {\bibinfo  {journal} {Nat.
  Phys.}\ }\textbf {\bibinfo {volume} {12}},\ \bibinfo {pages} {639} (\bibinfo
  {year} {2016})}\BibitemShut {NoStop}%
\bibitem [{\citenamefont {Madison}\ \emph {et~al.}(2000)\citenamefont
  {Madison}, \citenamefont {Chevy}, \citenamefont {Wohlleben},\ and\
  \citenamefont {Dalibard}}]{madison:2000}%
  \BibitemOpen
  \bibfield  {author} {\bibinfo {author} {\bibfnamefont {K.~W.}\ \bibnamefont
  {Madison}}, \bibinfo {author} {\bibfnamefont {F.}~\bibnamefont {Chevy}},
  \bibinfo {author} {\bibfnamefont {W.}~\bibnamefont {Wohlleben}}, \ and\
  \bibinfo {author} {\bibfnamefont {J.}~\bibnamefont {Dalibard}},\ }\href
  {\doibase 10.1103/PhysRevLett.84.806} {\bibfield  {journal} {\bibinfo
  {journal} {Phys. Rev. Lett.}\ }\textbf {\bibinfo {volume} {84}},\ \bibinfo
  {pages} {806} (\bibinfo {year} {2000})}\BibitemShut {NoStop}%
\bibitem [{\citenamefont {Abo-Shaeer}\ \emph {et~al.}(2001)\citenamefont
  {Abo-Shaeer}, \citenamefont {Raman}, \citenamefont {Vogels},\ and\
  \citenamefont {Ketterle}}]{abo-shaeer:2001}%
  \BibitemOpen
  \bibfield  {author} {\bibinfo {author} {\bibfnamefont {J.~R.}\ \bibnamefont
  {Abo-Shaeer}}, \bibinfo {author} {\bibfnamefont {C.}~\bibnamefont {Raman}},
  \bibinfo {author} {\bibfnamefont {J.~M.}\ \bibnamefont {Vogels}}, \ and\
  \bibinfo {author} {\bibfnamefont {W.}~\bibnamefont {Ketterle}},\ }\href
  {\doibase 10.1126/science.1060182} {\bibfield  {journal} {\bibinfo  {journal}
  {Science}\ }\textbf {\bibinfo {volume} {292}},\ \bibinfo {pages} {476}
  (\bibinfo {year} {2001})}\BibitemShut {NoStop}%
\bibitem [{\citenamefont {Bretin}\ \emph {et~al.}(2004)\citenamefont {Bretin},
  \citenamefont {Stock}, \citenamefont {Seurin},\ and\ \citenamefont
  {Dalibard}}]{bretin:2004}%
  \BibitemOpen
  \bibfield  {author} {\bibinfo {author} {\bibfnamefont {V.}~\bibnamefont
  {Bretin}}, \bibinfo {author} {\bibfnamefont {S.}~\bibnamefont {Stock}},
  \bibinfo {author} {\bibfnamefont {Y.}~\bibnamefont {Seurin}}, \ and\ \bibinfo
  {author} {\bibfnamefont {J.}~\bibnamefont {Dalibard}},\ }\href {\doibase
  10.1103/PhysRevLett.92.050403} {\bibfield  {journal} {\bibinfo  {journal}
  {Phys. Rev. Lett.}\ }\textbf {\bibinfo {volume} {92}},\ \bibinfo {pages}
  {050403} (\bibinfo {year} {2004})}\BibitemShut {NoStop}%
\bibitem [{\citenamefont {Schweikhard}\ \emph {et~al.}(2004)\citenamefont
  {Schweikhard}, \citenamefont {Coddington}, \citenamefont {Engels},
  \citenamefont {Mogendorff},\ and\ \citenamefont
  {Cornell}}]{schweikhard:2004}%
  \BibitemOpen
  \bibfield  {author} {\bibinfo {author} {\bibfnamefont {V.}~\bibnamefont
  {Schweikhard}}, \bibinfo {author} {\bibfnamefont {I.}~\bibnamefont
  {Coddington}}, \bibinfo {author} {\bibfnamefont {P.}~\bibnamefont {Engels}},
  \bibinfo {author} {\bibfnamefont {V.}~\bibnamefont {Mogendorff}}, \ and\
  \bibinfo {author} {\bibfnamefont {E.}~\bibnamefont {Cornell}},\ }\href
  {\doibase 10.1103/PhysRevLett.92.040404} {\bibfield  {journal} {\bibinfo
  {journal} {Phys. Rev. Lett.}\ }\textbf {\bibinfo {volume} {92}},\ \bibinfo
  {pages} {040404} (\bibinfo {year} {2004})}\BibitemShut {NoStop}%
\bibitem [{\citenamefont {Tung}\ \emph {et~al.}(2006)\citenamefont {Tung},
  \citenamefont {Schweikhard},\ and\ \citenamefont {Cornell}}]{tung:2006}%
  \BibitemOpen
  \bibfield  {author} {\bibinfo {author} {\bibfnamefont {S.}~\bibnamefont
  {Tung}}, \bibinfo {author} {\bibfnamefont {V.}~\bibnamefont {Schweikhard}}, \
  and\ \bibinfo {author} {\bibfnamefont {E.~A.}\ \bibnamefont {Cornell}},\
  }\href {\doibase 10.1103/PhysRevLett.97.240402} {\bibfield  {journal}
  {\bibinfo  {journal} {Phys. Rev. Lett.}\ }\textbf {\bibinfo {volume} {97}},\
  \bibinfo {pages} {240402} (\bibinfo {year} {2006})}\BibitemShut {NoStop}%
\bibitem [{\citenamefont {Gemelke}\ \emph {et~al.}(2010)\citenamefont
  {Gemelke}, \citenamefont {Sarajlic},\ and\ \citenamefont
  {Chu}}]{gemelke:2010}%
  \BibitemOpen
  \bibfield  {author} {\bibinfo {author} {\bibfnamefont {N.}~\bibnamefont
  {Gemelke}}, \bibinfo {author} {\bibfnamefont {E.}~\bibnamefont {Sarajlic}}, \
  and\ \bibinfo {author} {\bibfnamefont {S.}~\bibnamefont {Chu}},\ }\href
  {http://arxiv.org/abs/1007.2677} {\  (\bibinfo {year} {2010})},\ \Eprint
  {http://arxiv.org/abs/arxiv:1007.2677} {arxiv:1007.2677} \BibitemShut
  {NoStop}%
\bibitem [{\citenamefont {Lin}\ \emph {et~al.}(2009{\natexlab{a}})\citenamefont
  {Lin}, \citenamefont {Compton}, \citenamefont {Perry}, \citenamefont
  {Phillips}, \citenamefont {Porto},\ and\ \citenamefont
  {Spielman}}]{lin:2009}%
  \BibitemOpen
  \bibfield  {author} {\bibinfo {author} {\bibfnamefont {Y.~J.}\ \bibnamefont
  {Lin}}, \bibinfo {author} {\bibfnamefont {R.~L.}\ \bibnamefont {Compton}},
  \bibinfo {author} {\bibfnamefont {A.~R.}\ \bibnamefont {Perry}}, \bibinfo
  {author} {\bibfnamefont {W.~D.}\ \bibnamefont {Phillips}}, \bibinfo {author}
  {\bibfnamefont {J.~V.}\ \bibnamefont {Porto}}, \ and\ \bibinfo {author}
  {\bibfnamefont {I.~B.}\ \bibnamefont {Spielman}},\ }\href {\doibase
  10.1103/PhysRevLett.102.130401} {\bibfield  {journal} {\bibinfo  {journal}
  {Phys. Rev. Lett.}\ }\textbf {\bibinfo {volume} {102}},\ \bibinfo {pages}
  {130401} (\bibinfo {year} {2009}{\natexlab{a}})}\BibitemShut {NoStop}%
\bibitem [{\citenamefont {Lin}\ \emph {et~al.}(2009{\natexlab{b}})\citenamefont
  {Lin}, \citenamefont {Compton}, \citenamefont {Jimenez-Garcia}, \citenamefont
  {Porto},\ and\ \citenamefont {Spielman}}]{lin:2009a}%
  \BibitemOpen
  \bibfield  {author} {\bibinfo {author} {\bibfnamefont {Y.~J.}\ \bibnamefont
  {Lin}}, \bibinfo {author} {\bibfnamefont {R.~L.}\ \bibnamefont {Compton}},
  \bibinfo {author} {\bibfnamefont {K.}~\bibnamefont {Jimenez-Garcia}},
  \bibinfo {author} {\bibfnamefont {J.~V.}\ \bibnamefont {Porto}}, \ and\
  \bibinfo {author} {\bibfnamefont {I.~B.}\ \bibnamefont {Spielman}},\ }\href
  {\doibase 10.1038/nature08609} {\bibfield  {journal} {\bibinfo  {journal}
  {Nature}\ }\textbf {\bibinfo {volume} {462}},\ \bibinfo {pages} {628}
  (\bibinfo {year} {2009}{\natexlab{b}})}\BibitemShut {NoStop}%
\bibitem [{\citenamefont {Aidelsburger}\ \emph {et~al.}(2011)\citenamefont
  {Aidelsburger}, \citenamefont {Atala}, \citenamefont {Nascimb\`ene},
  \citenamefont {Trotzky}, \citenamefont {Chen},\ and\ \citenamefont
  {Bloch}}]{aidelsburger:2011}%
  \BibitemOpen
  \bibfield  {author} {\bibinfo {author} {\bibfnamefont {M.}~\bibnamefont
  {Aidelsburger}}, \bibinfo {author} {\bibfnamefont {M.}~\bibnamefont {Atala}},
  \bibinfo {author} {\bibfnamefont {S.}~\bibnamefont {Nascimb\`ene}}, \bibinfo
  {author} {\bibfnamefont {S.}~\bibnamefont {Trotzky}}, \bibinfo {author}
  {\bibfnamefont {Y.}~\bibnamefont {Chen}}, \ and\ \bibinfo {author}
  {\bibfnamefont {I.}~\bibnamefont {Bloch}},\ }\href {\doibase
  10.1103/PhysRevLett.107.255301} {\bibfield  {journal} {\bibinfo  {journal}
  {Phys. Rev. Lett.}\ }\textbf {\bibinfo {volume} {107}},\ \bibinfo {pages}
  {255301} (\bibinfo {year} {2011})}\BibitemShut {NoStop}%
\bibitem [{\citenamefont {LeBlanc}\ \emph {et~al.}(2012)\citenamefont
  {LeBlanc}, \citenamefont {Jimenez-Garcia}, \citenamefont {Williams},
  \citenamefont {Beeler}, \citenamefont {Perry}, \citenamefont {Phillips},\
  and\ \citenamefont {Spielman}}]{leblanc:2012}%
  \BibitemOpen
  \bibfield  {author} {\bibinfo {author} {\bibfnamefont {L.~J.}\ \bibnamefont
  {LeBlanc}}, \bibinfo {author} {\bibfnamefont {K.}~\bibnamefont
  {Jimenez-Garcia}}, \bibinfo {author} {\bibfnamefont {R.~A.}\ \bibnamefont
  {Williams}}, \bibinfo {author} {\bibfnamefont {M.~C.}\ \bibnamefont
  {Beeler}}, \bibinfo {author} {\bibfnamefont {A.~R.}\ \bibnamefont {Perry}},
  \bibinfo {author} {\bibfnamefont {W.~D.}\ \bibnamefont {Phillips}}, \ and\
  \bibinfo {author} {\bibfnamefont {I.~B.}\ \bibnamefont {Spielman}},\ }\href
  {\doibase 10.1073/pnas.1202579109} {\bibfield  {journal} {\bibinfo  {journal}
  {Proc. of the Nat. Acad. of Sci.}\ }\textbf {\bibinfo {volume} {109}},\
  \bibinfo {pages} {10811} (\bibinfo {year} {2012})}\BibitemShut {NoStop}%
\bibitem [{\citenamefont {Struck}\ \emph {et~al.}(2012)\citenamefont {Struck},
  \citenamefont {\"Olschl\"ager}, \citenamefont {Weinberg}, \citenamefont
  {Hauke}, \citenamefont {Simonet}, \citenamefont {Eckardt}, \citenamefont
  {Lewenstein}, \citenamefont {Sengstock},\ and\ \citenamefont
  {Windpassinger}}]{struck:2012}%
  \BibitemOpen
  \bibfield  {author} {\bibinfo {author} {\bibfnamefont {J.}~\bibnamefont
  {Struck}}, \bibinfo {author} {\bibfnamefont {C.}~\bibnamefont
  {\"Olschl\"ager}}, \bibinfo {author} {\bibfnamefont {M.}~\bibnamefont
  {Weinberg}}, \bibinfo {author} {\bibfnamefont {P.}~\bibnamefont {Hauke}},
  \bibinfo {author} {\bibfnamefont {J.}~\bibnamefont {Simonet}}, \bibinfo
  {author} {\bibfnamefont {A.}~\bibnamefont {Eckardt}}, \bibinfo {author}
  {\bibfnamefont {M.}~\bibnamefont {Lewenstein}}, \bibinfo {author}
  {\bibfnamefont {K.}~\bibnamefont {Sengstock}}, \ and\ \bibinfo {author}
  {\bibfnamefont {P.}~\bibnamefont {Windpassinger}},\ }\href {\doibase
  10.1103/PhysRevLett.108.225304} {\bibfield  {journal} {\bibinfo  {journal}
  {Phys. Rev. Lett.}\ }\textbf {\bibinfo {volume} {108}},\ \bibinfo {pages}
  {225304} (\bibinfo {year} {2012})}\BibitemShut {NoStop}%
\bibitem [{\citenamefont {Aidelsburger}\ \emph {et~al.}(2013)\citenamefont
  {Aidelsburger}, \citenamefont {Atala}, \citenamefont {Lohse}, \citenamefont
  {Barreiro}, \citenamefont {Paredes},\ and\ \citenamefont
  {Bloch}}]{aidelsburger:2013}%
  \BibitemOpen
  \bibfield  {author} {\bibinfo {author} {\bibfnamefont {M.}~\bibnamefont
  {Aidelsburger}}, \bibinfo {author} {\bibfnamefont {M.}~\bibnamefont {Atala}},
  \bibinfo {author} {\bibfnamefont {M.}~\bibnamefont {Lohse}}, \bibinfo
  {author} {\bibfnamefont {J.~T.}\ \bibnamefont {Barreiro}}, \bibinfo {author}
  {\bibfnamefont {B.}~\bibnamefont {Paredes}}, \ and\ \bibinfo {author}
  {\bibfnamefont {I.}~\bibnamefont {Bloch}},\ }\href
  {http://dx.doi.org/10.1103/PhysRevLett.111.185301} {\bibfield  {journal}
  {\bibinfo  {journal} {Phys. Rev. Lett.}\ }\textbf {\bibinfo {volume} {111}},\
  \bibinfo {pages} {185301} (\bibinfo {year} {2013})}\BibitemShut {NoStop}%
\bibitem [{\citenamefont {Miyake}\ \emph {et~al.}(2013)\citenamefont {Miyake},
  \citenamefont {Siviloglou}, \citenamefont {Kennedy}, \citenamefont {Burton},\
  and\ \citenamefont {Ketterle}}]{miyake:2013}%
  \BibitemOpen
  \bibfield  {author} {\bibinfo {author} {\bibfnamefont {H.}~\bibnamefont
  {Miyake}}, \bibinfo {author} {\bibfnamefont {G.~A.}\ \bibnamefont
  {Siviloglou}}, \bibinfo {author} {\bibfnamefont {C.~J.}\ \bibnamefont
  {Kennedy}}, \bibinfo {author} {\bibfnamefont {W.~C.}\ \bibnamefont {Burton}},
  \ and\ \bibinfo {author} {\bibfnamefont {W.}~\bibnamefont {Ketterle}},\
  }\href {http://dx.doi.org/10.1103/PhysRevLett.111.185302} {\bibfield
  {journal} {\bibinfo  {journal} {Phys. Rev. Lett.}\ }\textbf {\bibinfo
  {volume} {111}},\ \bibinfo {pages} {185302} (\bibinfo {year}
  {2013})}\BibitemShut {NoStop}%
\bibitem [{\citenamefont {Atala}\ \emph {et~al.}(2014)\citenamefont {Atala},
  \citenamefont {Aidelsburger}, \citenamefont {Lohse}, \citenamefont
  {Barreiro}, \citenamefont {Paredes},\ and\ \citenamefont
  {Bloch}}]{atala:2014}%
  \BibitemOpen
  \bibfield  {author} {\bibinfo {author} {\bibfnamefont {M.}~\bibnamefont
  {Atala}}, \bibinfo {author} {\bibfnamefont {M.}~\bibnamefont {Aidelsburger}},
  \bibinfo {author} {\bibfnamefont {M.}~\bibnamefont {Lohse}}, \bibinfo
  {author} {\bibfnamefont {J.~T.}\ \bibnamefont {Barreiro}}, \bibinfo {author}
  {\bibfnamefont {B.}~\bibnamefont {Paredes}}, \ and\ \bibinfo {author}
  {\bibfnamefont {I.}~\bibnamefont {Bloch}},\ }\href {\doibase
  10.1038/nphys2998} {\bibfield  {journal} {\bibinfo  {journal} {Nat. Phys.}\
  }\textbf {\bibinfo {volume} {10}},\ \bibinfo {pages} {588} (\bibinfo {year}
  {2014})}\BibitemShut {NoStop}%
\bibitem [{\citenamefont {Stuhl}\ \emph {et~al.}(2015)\citenamefont {Stuhl},
  \citenamefont {Lu}, \citenamefont {Aycock}, \citenamefont {Genkina},\ and\
  \citenamefont {Spielman}}]{stuhl:2015}%
  \BibitemOpen
  \bibfield  {author} {\bibinfo {author} {\bibfnamefont {B.~K.}\ \bibnamefont
  {Stuhl}}, \bibinfo {author} {\bibfnamefont {H.}~\bibnamefont {Lu}}, \bibinfo
  {author} {\bibfnamefont {L.~M.}\ \bibnamefont {Aycock}}, \bibinfo {author}
  {\bibfnamefont {D.}~\bibnamefont {Genkina}}, \ and\ \bibinfo {author}
  {\bibfnamefont {I.~B.}\ \bibnamefont {Spielman}},\ }\href {\doibase
  10.1126/science.aaa8515} {\bibfield  {journal} {\bibinfo  {journal}
  {Science}\ }\textbf {\bibinfo {volume} {349}},\ \bibinfo {pages} {1514}
  (\bibinfo {year} {2015})}\BibitemShut {NoStop}%
\bibitem [{\citenamefont {Kennedy}\ \emph {et~al.}(2015)\citenamefont
  {Kennedy}, \citenamefont {Burton}, \citenamefont {Chung},\ and\ \citenamefont
  {Ketterle}}]{kennedy:2015}%
  \BibitemOpen
  \bibfield  {author} {\bibinfo {author} {\bibfnamefont {C.~J.}\ \bibnamefont
  {Kennedy}}, \bibinfo {author} {\bibfnamefont {W.~C.}\ \bibnamefont {Burton}},
  \bibinfo {author} {\bibfnamefont {W.~C.}\ \bibnamefont {Chung}}, \ and\
  \bibinfo {author} {\bibfnamefont {W.}~\bibnamefont {Ketterle}},\ }\href
  {\doibase 10.1038/nphys3421} {\bibfield  {journal} {\bibinfo  {journal} {Nat.
  Phys.}\ }\textbf {\bibinfo {volume} {11}},\ \bibinfo {pages} {859} (\bibinfo
  {year} {2015})}\BibitemShut {NoStop}%
\bibitem [{\citenamefont {Scarola}\ and\ \citenamefont
  {Das~Sarma}(2007)}]{scarola:2007}%
  \BibitemOpen
  \bibfield  {author} {\bibinfo {author} {\bibfnamefont {V.~W.}\ \bibnamefont
  {Scarola}}\ and\ \bibinfo {author} {\bibfnamefont {S.}~\bibnamefont
  {Das~Sarma}},\ }\href {\doibase 10.1103/PhysRevLett.98.210403} {\bibfield
  {journal} {\bibinfo  {journal} {Phys. Rev. Lett.}\ }\textbf {\bibinfo
  {volume} {98}},\ \bibinfo {pages} {210403} (\bibinfo {year}
  {2007})}\BibitemShut {NoStop}%
\bibitem [{\citenamefont {Zhao}\ \emph {et~al.}(2011)\citenamefont {Zhao},
  \citenamefont {Bray-Ali}, \citenamefont {Williams}, \citenamefont
  {Spielman},\ and\ \citenamefont {Satija}}]{zhao:2011}%
  \BibitemOpen
  \bibfield  {author} {\bibinfo {author} {\bibfnamefont {E.}~\bibnamefont
  {Zhao}}, \bibinfo {author} {\bibfnamefont {N.}~\bibnamefont {Bray-Ali}},
  \bibinfo {author} {\bibfnamefont {C.~J.}\ \bibnamefont {Williams}}, \bibinfo
  {author} {\bibfnamefont {I.~B.}\ \bibnamefont {Spielman}}, \ and\ \bibinfo
  {author} {\bibfnamefont {I.~I.}\ \bibnamefont {Satija}},\ }\href {\doibase
  10.1103/PhysRevA.84.063629} {\bibfield  {journal} {\bibinfo  {journal} {Phys.
  Rev. A}\ }\textbf {\bibinfo {volume} {84}},\ \bibinfo {pages} {063629}
  (\bibinfo {year} {2011})}\BibitemShut {NoStop}%
\bibitem [{\citenamefont {{Goldman}}\ \emph {et~al.}(2013)\citenamefont
  {{Goldman}}, \citenamefont {{Juzeliunas}}, \citenamefont {{Ohberg}},\ and\
  \citenamefont {{Spielman}}}]{goldman:2013}%
  \BibitemOpen
  \bibfield  {author} {\bibinfo {author} {\bibfnamefont {N.}~\bibnamefont
  {{Goldman}}}, \bibinfo {author} {\bibfnamefont {G.}~\bibnamefont
  {{Juzeliunas}}}, \bibinfo {author} {\bibfnamefont {P.}~\bibnamefont
  {{Ohberg}}}, \ and\ \bibinfo {author} {\bibfnamefont {I.~B.}\ \bibnamefont
  {{Spielman}}},\ }\href {http://adsabs.harvard.edu/abs/2013arXiv1308.6533G}
  {\bibfield  {journal} {\bibinfo  {journal} {ArXiv e-prints}\ } (\bibinfo
  {year} {2013})},\ \Eprint {http://arxiv.org/abs/1308.6533} {arXiv:1308.6533}
  \BibitemShut {NoStop}%
\bibitem [{\citenamefont {Sarma}\ and\ \citenamefont
  {Pinczuk}(1997)}]{sarma:1997}%
  \BibitemOpen
  \bibinfo {editor} {\bibfnamefont {S.~D.}\ \bibnamefont {Sarma}}\ and\
  \bibinfo {editor} {\bibfnamefont {A.}~\bibnamefont {Pinczuk}},\ eds.,\
  \href@noop {} {\emph {\bibinfo {title} {Perspectives in quantum {H}all
  effects: {N}ovel quantum liquids in low-dimensional semiconductor
  structures}}}\ (\bibinfo  {publisher} {Wiley},\ \bibinfo {address} {New
  York},\ \bibinfo {year} {1997})\BibitemShut {NoStop}%
\bibitem [{sup()}]{supplmentaryinfo}%
  \BibitemOpen
  \href@noop {} {}\bibinfo {note} {See Supplemental Material}\BibitemShut
  {NoStop}%
\bibitem [{\citenamefont {Yoshioka}(1987)}]{yoshioka:1987}%
  \BibitemOpen
  \bibfield  {author} {\bibinfo {author} {\bibfnamefont {D.}~\bibnamefont
  {Yoshioka}},\ }\href {\doibase 10.1143/jpsj.56.1301} {\bibfield  {journal}
  {\bibinfo  {journal} {J. Phys. Soc. Jpn.}\ }\textbf {\bibinfo {volume}
  {56}},\ \bibinfo {pages} {1301} (\bibinfo {year} {1987})}\BibitemShut
  {NoStop}%
\bibitem [{\citenamefont {Chakraborty}\ and\ \citenamefont
  {Pietil{\"a}inen}(1997)}]{chakraborty:1997}%
  \BibitemOpen
  \bibfield  {author} {\bibinfo {author} {\bibfnamefont {T.}~\bibnamefont
  {Chakraborty}}\ and\ \bibinfo {author} {\bibfnamefont {P.}~\bibnamefont
  {Pietil{\"a}inen}},\ }\href {\doibase 10.1103/PhysRevB.55.R1954} {\bibfield
  {journal} {\bibinfo  {journal} {Phys. Rev. B}\ }\textbf {\bibinfo {volume}
  {55}},\ \bibinfo {pages} {R1954} (\bibinfo {year} {1997})}\BibitemShut
  {NoStop}%
\bibitem [{\citenamefont {Zheng}\ and\ \citenamefont
  {MacDonald}(1994)}]{zheng:1994}%
  \BibitemOpen
  \bibfield  {author} {\bibinfo {author} {\bibfnamefont {L.}~\bibnamefont
  {Zheng}}\ and\ \bibinfo {author} {\bibfnamefont {A.}~\bibnamefont
  {MacDonald}},\ }\href {\doibase
  http://dx.doi.org/10.1016/0039-6028(94)90867-2} {\bibfield  {journal}
  {\bibinfo  {journal} {Surf. Sci.}\ }\textbf {\bibinfo {volume} {305}},\
  \bibinfo {pages} {101} (\bibinfo {year} {1994})}\BibitemShut {NoStop}%
\bibitem [{\citenamefont {Tevosyan}\ and\ \citenamefont
  {MacDonald}(1997)}]{tevosyan:1997}%
  \BibitemOpen
  \bibfield  {author} {\bibinfo {author} {\bibfnamefont {K.}~\bibnamefont
  {Tevosyan}}\ and\ \bibinfo {author} {\bibfnamefont {A.~H.}\ \bibnamefont
  {MacDonald}},\ }\href {\doibase 10.1103/PhysRevB.56.7517} {\bibfield
  {journal} {\bibinfo  {journal} {Phys. Rev. B}\ }\textbf {\bibinfo {volume}
  {56}},\ \bibinfo {pages} {7517} (\bibinfo {year} {1997})}\BibitemShut
  {NoStop}%
\bibitem [{\citenamefont {Sawatdiaree}\ and\ \citenamefont
  {Apel}(2000)}]{sawatdiaree:2000}%
  \BibitemOpen
  \bibfield  {author} {\bibinfo {author} {\bibfnamefont {S.}~\bibnamefont
  {Sawatdiaree}}\ and\ \bibinfo {author} {\bibfnamefont {W.}~\bibnamefont
  {Apel}},\ }\href {\doibase http://dx.doi.org/10.1016/S1386-9477(99)00062-4}
  {\bibfield  {journal} {\bibinfo  {journal} {Physica E}\ }\textbf {\bibinfo
  {volume} {6}},\ \bibinfo {pages} {75} (\bibinfo {year} {2000})}\BibitemShut
  {NoStop}%
\bibitem [{\citenamefont {Murthy}\ and\ \citenamefont
  {Shankar}(2003)}]{murthy:2003}%
  \BibitemOpen
  \bibfield  {author} {\bibinfo {author} {\bibfnamefont {G.}~\bibnamefont
  {Murthy}}\ and\ \bibinfo {author} {\bibfnamefont {R.}~\bibnamefont
  {Shankar}},\ }\href {\doibase 10.1103/RevModPhys.75.1101} {\bibfield
  {journal} {\bibinfo  {journal} {Rev. Mod. Phys.}\ }\textbf {\bibinfo {volume}
  {75}},\ \bibinfo {pages} {1101} (\bibinfo {year} {2003})}\BibitemShut
  {NoStop}%
\bibitem [{\citenamefont {Jain}(1989)}]{jain:1989}%
  \BibitemOpen
  \bibfield  {author} {\bibinfo {author} {\bibfnamefont {J.}~\bibnamefont
  {Jain}},\ }\href {http://dx.doi.org/10.1103/PhysRevLett.63.199} {\bibfield
  {journal} {\bibinfo  {journal} {Phys. Rev. Lett.}\ }\textbf {\bibinfo
  {volume} {63}},\ \bibinfo {pages} {199} (\bibinfo {year} {1989})}\BibitemShut
  {NoStop}%
\bibitem [{\citenamefont {Jain}(2007)}]{jain:2007}%
  \BibitemOpen
  \bibfield  {author} {\bibinfo {author} {\bibfnamefont {J.}~\bibnamefont
  {Jain}},\ }\href@noop {} {\emph {\bibinfo {title} {Composite fermions}}}\
  (\bibinfo  {publisher} {Cambridge University Press},\ \bibinfo {year}
  {2007})\BibitemShut {NoStop}%
\bibitem [{\citenamefont {Jain}\ \emph {et~al.}(2005)\citenamefont {Jain},
  \citenamefont {Park}, \citenamefont {Peterson},\ and\ \citenamefont
  {Scarola}}]{jain:2005}%
  \BibitemOpen
  \bibfield  {author} {\bibinfo {author} {\bibfnamefont {J.}~\bibnamefont
  {Jain}}, \bibinfo {author} {\bibfnamefont {K.}~\bibnamefont {Park}}, \bibinfo
  {author} {\bibfnamefont {M.}~\bibnamefont {Peterson}}, \ and\ \bibinfo
  {author} {\bibfnamefont {V.}~\bibnamefont {Scarola}},\ }\href {\doibase
  10.1016/j.sse.2005.04.033} {\bibfield  {journal} {\bibinfo  {journal} {Solid
  State Commun.}\ }\textbf {\bibinfo {volume} {135}},\ \bibinfo {pages} {602}
  (\bibinfo {year} {2005})}\BibitemShut {NoStop}%
\bibitem [{\citenamefont {Weisse}\ \emph {et~al.}(2006)\citenamefont {Weisse},
  \citenamefont {Wellein}, \citenamefont {Alvermann},\ and\ \citenamefont
  {Fehske}}]{weisse:2006}%
  \BibitemOpen
  \bibfield  {author} {\bibinfo {author} {\bibfnamefont {A.}~\bibnamefont
  {Weisse}}, \bibinfo {author} {\bibfnamefont {G.}~\bibnamefont {Wellein}},
  \bibinfo {author} {\bibfnamefont {A.}~\bibnamefont {Alvermann}}, \ and\
  \bibinfo {author} {\bibfnamefont {H.}~\bibnamefont {Fehske}},\ }\href
  {\doibase 10.1103/RevModPhys.78.275} {\bibfield  {journal} {\bibinfo
  {journal} {Rev. Mod. Phys.}\ }\textbf {\bibinfo {volume} {78}},\ \bibinfo
  {pages} {275} (\bibinfo {year} {2006})}\BibitemShut {NoStop}%
\bibitem [{\citenamefont {Haldane}(1983)}]{haldane:1983}%
  \BibitemOpen
  \bibfield  {author} {\bibinfo {author} {\bibfnamefont {F.}~\bibnamefont
  {Haldane}},\ }\href {http://dx.doi.org/10.1103/PhysRevLett.51.605} {\bibfield
   {journal} {\bibinfo  {journal} {Phys. Rev. Lett.}\ }\textbf {\bibinfo
  {volume} {51}},\ \bibinfo {pages} {605} (\bibinfo {year} {1983})}\BibitemShut
  {NoStop}%
\bibitem [{\citenamefont {Trugman}\ and\ \citenamefont
  {Kivelson}(1985)}]{trugman:1985}%
  \BibitemOpen
  \bibfield  {author} {\bibinfo {author} {\bibfnamefont {S.}~\bibnamefont
  {Trugman}}\ and\ \bibinfo {author} {\bibfnamefont {S.}~\bibnamefont
  {Kivelson}},\ }\href {http://dx.doi.org/10.1103/PhysRevB.31.5280} {\bibfield
  {journal} {\bibinfo  {journal} {Phys. Rev. B}\ }\textbf {\bibinfo {volume}
  {31}},\ \bibinfo {pages} {5280} (\bibinfo {year} {1985})}\BibitemShut
  {NoStop}%
\bibitem [{\citenamefont {Mottelson}(1999)}]{mottelson:1999}%
  \BibitemOpen
  \bibfield  {author} {\bibinfo {author} {\bibfnamefont {B.}~\bibnamefont
  {Mottelson}},\ }\href {http://dx.doi.org/10.1103/PhysRevLett.83.2695}
  {\bibfield  {journal} {\bibinfo  {journal} {Phys. Rev. Lett.}\ }\textbf
  {\bibinfo {volume} {83}},\ \bibinfo {pages} {2695} (\bibinfo {year}
  {1999})}\BibitemShut {NoStop}%
\bibitem [{\citenamefont {Cooper}\ and\ \citenamefont
  {Wilkin}(1999)}]{cooper:1999}%
  \BibitemOpen
  \bibfield  {author} {\bibinfo {author} {\bibfnamefont {N.~R.}\ \bibnamefont
  {Cooper}}\ and\ \bibinfo {author} {\bibfnamefont {N.~K.}\ \bibnamefont
  {Wilkin}},\ }\href {http://link.aps.org/doi/10.1103/PhysRevB.60.R16279}
  {\bibfield  {journal} {\bibinfo  {journal} {Phys. Rev. B}\ }\textbf {\bibinfo
  {volume} {60}},\ \bibinfo {pages} {R16279} (\bibinfo {year}
  {1999})}\BibitemShut {NoStop}%
\bibitem [{\citenamefont {Wilkin}\ and\ \citenamefont
  {Gunn}(2000)}]{wilkin:2000}%
  \BibitemOpen
  \bibfield  {author} {\bibinfo {author} {\bibfnamefont {N.}~\bibnamefont
  {Wilkin}}\ and\ \bibinfo {author} {\bibfnamefont {J.}~\bibnamefont {Gunn}},\
  }\href {http://dx.doi.org/10.1103/PhysRevLett.84.6} {\bibfield  {journal}
  {\bibinfo  {journal} {Phys. Rev. Lett.}\ }\textbf {\bibinfo {volume} {84}},\
  \bibinfo {pages} {6} (\bibinfo {year} {2000})}\BibitemShut {NoStop}%
\bibitem [{\citenamefont {Viefers}\ \emph {et~al.}(2000)\citenamefont
  {Viefers}, \citenamefont {Hansson},\ and\ \citenamefont
  {Reimann}}]{viefers:2000}%
  \BibitemOpen
  \bibfield  {author} {\bibinfo {author} {\bibfnamefont {S.}~\bibnamefont
  {Viefers}}, \bibinfo {author} {\bibfnamefont {T.~H.}\ \bibnamefont
  {Hansson}}, \ and\ \bibinfo {author} {\bibfnamefont {S.~M.}\ \bibnamefont
  {Reimann}},\ }\href {http://link.aps.org/doi/10.1103/PhysRevA.62.053604}
  {\bibfield  {journal} {\bibinfo  {journal} {Phys. Rev. A}\ }\textbf {\bibinfo
  {volume} {62}},\ \bibinfo {pages} {053604} (\bibinfo {year}
  {2000})}\BibitemShut {NoStop}%
\bibitem [{\citenamefont {Regnault}\ and\ \citenamefont
  {Jolicoeur}(2003)}]{regnault:2003}%
  \BibitemOpen
  \bibfield  {author} {\bibinfo {author} {\bibfnamefont {N.}~\bibnamefont
  {Regnault}}\ and\ \bibinfo {author} {\bibfnamefont {T.}~\bibnamefont
  {Jolicoeur}},\ }\href {\doibase 10.1103/PhysRevLett.91.030402} {\bibfield
  {journal} {\bibinfo  {journal} {Phys. Rev. Lett.}\ }\textbf {\bibinfo
  {volume} {91}},\ \bibinfo {pages} {030402} (\bibinfo {year}
  {2003})}\BibitemShut {NoStop}%
\bibitem [{\citenamefont {Regnault}\ and\ \citenamefont
  {Jolicoeur}(2004)}]{regnault:2004}%
  \BibitemOpen
  \bibfield  {author} {\bibinfo {author} {\bibfnamefont {N.}~\bibnamefont
  {Regnault}}\ and\ \bibinfo {author} {\bibfnamefont {T.}~\bibnamefont
  {Jolicoeur}},\ }\href {http://link.aps.org/doi/10.1103/PhysRevB.69.235309}
  {\bibfield  {journal} {\bibinfo  {journal} {Phys. Rev. B}\ }\textbf {\bibinfo
  {volume} {69}},\ \bibinfo {pages} {235309} (\bibinfo {year}
  {2004})}\BibitemShut {NoStop}%
\bibitem [{\citenamefont {Sorensen}\ \emph {et~al.}(2005)\citenamefont
  {Sorensen}, \citenamefont {Demler},\ and\ \citenamefont
  {Lukin}}]{sorensen:2005}%
  \BibitemOpen
  \bibfield  {author} {\bibinfo {author} {\bibfnamefont {A.}~\bibnamefont
  {Sorensen}}, \bibinfo {author} {\bibfnamefont {E.}~\bibnamefont {Demler}}, \
  and\ \bibinfo {author} {\bibfnamefont {M.}~\bibnamefont {Lukin}},\ }\href
  {\doibase 10.1103/PhysRevLett.94.086803} {\bibfield  {journal} {\bibinfo
  {journal} {Phys. Rev. Lett.}\ }\textbf {\bibinfo {volume} {94}},\ \bibinfo
  {pages} {086803} (\bibinfo {year} {2005})}\BibitemShut {NoStop}%
\bibitem [{\citenamefont {Fano}\ \emph {et~al.}(1986)\citenamefont {Fano},
  \citenamefont {Ortolani},\ and\ \citenamefont {Colombo}}]{fano:1986}%
  \BibitemOpen
  \bibfield  {author} {\bibinfo {author} {\bibfnamefont {G.}~\bibnamefont
  {Fano}}, \bibinfo {author} {\bibfnamefont {F.}~\bibnamefont {Ortolani}}, \
  and\ \bibinfo {author} {\bibfnamefont {E.}~\bibnamefont {Colombo}},\ }\href
  {\doibase 10.1103/PhysRevB.34.2670} {\bibfield  {journal} {\bibinfo
  {journal} {Phys. Rev. B}\ }\textbf {\bibinfo {volume} {34}},\ \bibinfo
  {pages} {2670} (\bibinfo {year} {1986})}\BibitemShut {NoStop}%
\bibitem [{\citenamefont {W{\'o}js}\ and\ \citenamefont
  {Quinn}(1998)}]{wojs:1998}%
  \BibitemOpen
  \bibfield  {author} {\bibinfo {author} {\bibfnamefont {A.}~\bibnamefont
  {W{\'o}js}}\ and\ \bibinfo {author} {\bibfnamefont {J.~J.}\ \bibnamefont
  {Quinn}},\ }\href
  {http://www.sciencedirect.com/science/article/pii/S1386947798002367}
  {\bibfield  {journal} {\bibinfo  {journal} {Physica E}\ }\textbf {\bibinfo
  {volume} {3}},\ \bibinfo {pages} {181} (\bibinfo {year} {1998})}\BibitemShut
  {NoStop}%
\bibitem [{\citenamefont {Wilkin}\ \emph {et~al.}(1998)\citenamefont {Wilkin},
  \citenamefont {Gunn},\ and\ \citenamefont {Smith}}]{wilkin:1998}%
  \BibitemOpen
  \bibfield  {author} {\bibinfo {author} {\bibfnamefont {N.~K.}\ \bibnamefont
  {Wilkin}}, \bibinfo {author} {\bibfnamefont {J.~M.~F.}\ \bibnamefont {Gunn}},
  \ and\ \bibinfo {author} {\bibfnamefont {R.~A.}\ \bibnamefont {Smith}},\
  }\href {\doibase 10.1103/PhysRevLett.80.2265} {\bibfield  {journal} {\bibinfo
   {journal} {Phys. Rev. Lett.}\ }\textbf {\bibinfo {volume} {80}},\ \bibinfo
  {pages} {2265} (\bibinfo {year} {1998})}\BibitemShut {NoStop}%
\bibitem [{\citenamefont {Nakajima}\ and\ \citenamefont
  {Ueda}(2003)}]{nakajima:2003}%
  \BibitemOpen
  \bibfield  {author} {\bibinfo {author} {\bibfnamefont {T.}~\bibnamefont
  {Nakajima}}\ and\ \bibinfo {author} {\bibfnamefont {M.}~\bibnamefont
  {Ueda}},\ }\href {\doibase 10.1103/PhysRevLett.91.140401} {\bibfield
  {journal} {\bibinfo  {journal} {Phys. Rev. Lett.}\ }\textbf {\bibinfo
  {volume} {91}},\ \bibinfo {pages} {140401} (\bibinfo {year}
  {2003})}\BibitemShut {NoStop}%
\bibitem [{\citenamefont {Rezayi}\ \emph {et~al.}(2005)\citenamefont {Rezayi},
  \citenamefont {Read},\ and\ \citenamefont {Cooper}}]{rezayi:2005}%
  \BibitemOpen
  \bibfield  {author} {\bibinfo {author} {\bibfnamefont {E.}~\bibnamefont
  {Rezayi}}, \bibinfo {author} {\bibfnamefont {N.}~\bibnamefont {Read}}, \ and\
  \bibinfo {author} {\bibfnamefont {N.}~\bibnamefont {Cooper}},\ }\href
  {\doibase 10.1103/PhysRevLett.95.160404} {\bibfield  {journal} {\bibinfo
  {journal} {Phys. Rev. Lett.}\ }\textbf {\bibinfo {volume} {95}},\ \bibinfo
  {pages} {160404} (\bibinfo {year} {2005})}\BibitemShut {NoStop}%
\bibitem [{\citenamefont {Osterloh}\ \emph {et~al.}(2007)\citenamefont
  {Osterloh}, \citenamefont {Barber{\'{a}}n},\ and\ \citenamefont
  {Lewenstein}}]{osterloh:2007}%
  \BibitemOpen
  \bibfield  {author} {\bibinfo {author} {\bibfnamefont {K.}~\bibnamefont
  {Osterloh}}, \bibinfo {author} {\bibfnamefont {N.}~\bibnamefont
  {Barber{\'{a}}n}}, \ and\ \bibinfo {author} {\bibfnamefont {M.}~\bibnamefont
  {Lewenstein}},\ }\href {http://dx.doi.org/10.1103/PhysRevLett.99.160403}
  {\bibfield  {journal} {\bibinfo  {journal} {Phys. Rev. Lett.}\ }\textbf
  {\bibinfo {volume} {99}} (\bibinfo {year} {2007})}\BibitemShut {NoStop}%
\bibitem [{\citenamefont {Grusdt}\ and\ \citenamefont
  {Fleischhauer}(2013)}]{grusdt:2013}%
  \BibitemOpen
  \bibfield  {author} {\bibinfo {author} {\bibfnamefont {F.}~\bibnamefont
  {Grusdt}}\ and\ \bibinfo {author} {\bibfnamefont {M.}~\bibnamefont
  {Fleischhauer}},\ }\href {http://dx.doi.org/10.1103/PhysRevA.87.043628}
  {\bibfield  {journal} {\bibinfo  {journal} {Phys. Rev. A}\ }\textbf {\bibinfo
  {volume} {87}} (\bibinfo {year} {2013})}\BibitemShut {NoStop}%
\bibitem [{\citenamefont {Gervais}\ and\ \citenamefont
  {Yang}(2010)}]{gervais:2010}%
  \BibitemOpen
  \bibfield  {author} {\bibinfo {author} {\bibfnamefont {G.}~\bibnamefont
  {Gervais}}\ and\ \bibinfo {author} {\bibfnamefont {K.}~\bibnamefont {Yang}},\
  }\href {\doibase 10.1103/PhysRevLett.105.086801} {\bibfield  {journal}
  {\bibinfo  {journal} {Phys. Rev. Lett.}\ }\textbf {\bibinfo {volume} {105}},\
  \bibinfo {pages} {086801} (\bibinfo {year} {2010})}\BibitemShut {NoStop}%
\bibitem [{\citenamefont {Kalmeyer}\ and\ \citenamefont
  {Laughlin}(1987)}]{kalmeyer:1987}%
  \BibitemOpen
  \bibfield  {author} {\bibinfo {author} {\bibfnamefont {V.}~\bibnamefont
  {Kalmeyer}}\ and\ \bibinfo {author} {\bibfnamefont {R.}~\bibnamefont
  {Laughlin}},\ }\href {http://dx.doi.org/10.1103/PhysRevLett.59.2095}
  {\bibfield  {journal} {\bibinfo  {journal} {Phys. Rev. Lett.}\ }\textbf
  {\bibinfo {volume} {59}},\ \bibinfo {pages} {2095} (\bibinfo {year}
  {1987})}\BibitemShut {NoStop}%
\bibitem [{\citenamefont {Yao}\ \emph {et~al.}(2013)\citenamefont {Yao},
  \citenamefont {Gorshkov}, \citenamefont {Laumann}, \citenamefont
  {L{\"a}uchli}, \citenamefont {Ye},\ and\ \citenamefont {Lukin}}]{yao:2013}%
  \BibitemOpen
  \bibfield  {author} {\bibinfo {author} {\bibfnamefont {N.~Y.}\ \bibnamefont
  {Yao}}, \bibinfo {author} {\bibfnamefont {A.~V.}\ \bibnamefont {Gorshkov}},
  \bibinfo {author} {\bibfnamefont {C.~R.}\ \bibnamefont {Laumann}}, \bibinfo
  {author} {\bibfnamefont {A.~M.}\ \bibnamefont {L{\"a}uchli}}, \bibinfo
  {author} {\bibfnamefont {J.}~\bibnamefont {Ye}}, \ and\ \bibinfo {author}
  {\bibfnamefont {M.~D.}\ \bibnamefont {Lukin}},\ }\href {\doibase
  10.1103/PhysRevLett.110.185302} {\bibfield  {journal} {\bibinfo  {journal}
  {Phys. Rev. Lett.}\ }\textbf {\bibinfo {volume} {110}},\ \bibinfo {pages}
  {185302} (\bibinfo {year} {2013})}\BibitemShut {NoStop}%
\end{thebibliography}%

\section{Supplementary Material for ``Thermometry for  Laughlin States of Ultracold Atoms''}

\subsection{Non-interacting Model of Excitations}

At temperatures well below the gap we understand the thermodynamics of the FQH regime in a na\"{i}ve two-level approximation.  We consider $N$ non-interacting particles with a gap, $\Delta$, to a set of $N$ degenerate modes (as in non-interacting CF theory).  With $k_B \equiv 1$, the partition function becomes:

\begin{eqnarray}
 Z_N = (1 + e^{-\Delta/T} )^N.
 \label{eq_ZN}
\end{eqnarray}
The entropy per particle obtained from $s_0=N^{-1}\partial (T \log Z_N )/\partial T$ becomes:

\begin{eqnarray}
s_0=\frac{\Delta}{T(1 + e^{-\Delta/T} )}e^{-\Delta/T}+\log{(1 + e^{-\Delta/T})}.
\end{eqnarray}
The heat capacity per particle obtained from $c_0=T\partial s_0/\partial T$ is then:
\begin{eqnarray}
c_0 = \left [\frac{\Delta}{T(1 + e^{-\Delta/T} )} \right ]^2e^{-\Delta/T}.  
\end{eqnarray}
This form for $c_0$ is a reasonable approximation for gapped states in the FQH regime for $T\ll \Delta$. 

At intermediate temperatures, $c_0$ reveals the temperature of the Schottky peak, $T_p$, obtained by setting $dc_0/dT\vert_{T=T_p}=0$.    $T_p$ separates exponential temperature dependence (for $T\ll T_p$) from power law temperature dependence (for $T\gg T_p$) in the heat capacity.  
Within this two-level approximation we find the corresponding entropy per particle at $T_p$ to be $s_0\vert_{T=T_p}\approx 0.29$. This estimate shows that one must lower the entropy per particle below $0.29$ to observe exponential temperature dependence in the heat capacity in a non-interacting gapped system.  

The dashed line in the bottom panel of Fig. 3 in the main text shows that $s_0$ starts to depart from the ansatz (solid line) well before $T_L$.  Here we expect $s_0$ to be inaccurate (in comparison to exact results and therefore the ansatz) because $s_0$ excludes the continuum and other aspects of the excited state structure.  $s_0$ could be slightly improved by noting that 
Eq.~\eqref{eq_ZN} is consistent with a theory of just non-interacting CFs at low energies because it just assumes a gap to $N$ degenerate modes.  A more sophisticated CF-only model would include inter-CF interactions to better capture the exact excitation state space.  Inclusion of inter-CF interactions in an improved version of Eq.~\eqref{eq_ZN} would bring the dashed line in the bottom panel of Fig.~3 of the main text into better agreement with the solid line at low $T$.  But as $T$ nears $T_L$ a method for including a more general set of excitations, i.e., the continuum, must be included.

\subsection{Stochastic Trace Method}

This section outlines a method to numerically compute the cumulants.  We first note that the cumulants are related to the moments of the Hamiltonian, $\text{Tr}(H^l)$, via:
\begin{eqnarray}
\kappa_l=\text{Tr}(H^l)-\sum_{l'=1}^{l-1} {l-1 \choose l'-1} \kappa_{l'} \text{Tr}(H^{l-l'}), 
\end{eqnarray}
for $l>0$. This expression allows us to estimate the cumulants with moments.  Consider $R$ normalized random vectors $\vert r\rangle$ in the Hilbert space.  The moments can then be estimated using:
\begin{eqnarray}
\text{Tr}(H^l)\approx R^{-1}\sum_{r=0}^{R-1} \langle r \vert H^l \vert r \rangle.
\end{eqnarray}
 We obtain convergence in the sum by choosing $R=100$ random vectors.  The error in this method scales like $\mathcal{O}[(RD)^{-1/2}]$ where, $D$ is the size of the Hilbert space.  For $N=12$ we achieve errors of less than $\sim10^{-6}$ and have compared with exact diagonalization where possible.  We therefore find that the stochastic trace method is numerically exact. 
 
\subsection{Finite-Size Scaling}

\begin{figure}[t]
	\centering
	\includegraphics[scale =.5]{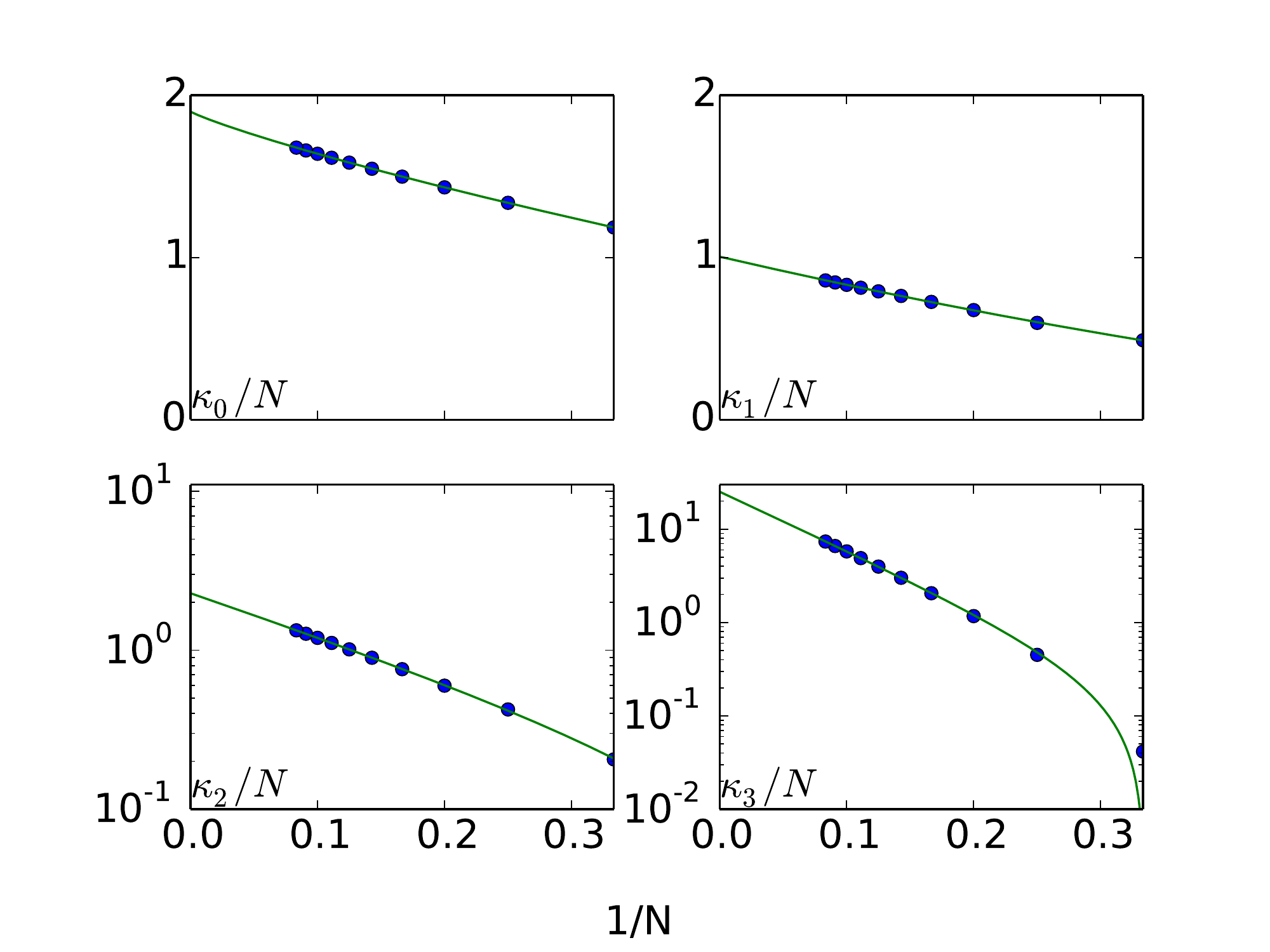}
	\caption {The symbols are boson cumulants per particle obtained from the numerically exact stochastic trace method.  The lines show finite size scaling functions Eqs.~\eqref{eq_s_kappa0}, \eqref{eq_s_kappa1}, and \eqref{eq_s_kappag}.  $\kappa_0$ is the log of the Hilbert space size and exhibits a log-linear $N$ scaling. $\kappa_1$ fits to a polynomial while $\kappa_{l>1}$ fit to an exponential.   The thermodynamic limit ($N\rightarrow\infty$) of these vales is given in Table~I in the main text.}
	\label{fig_cumulants}
\end{figure}

\begin{figure}[t]
	\centering
	\includegraphics[scale=.45]{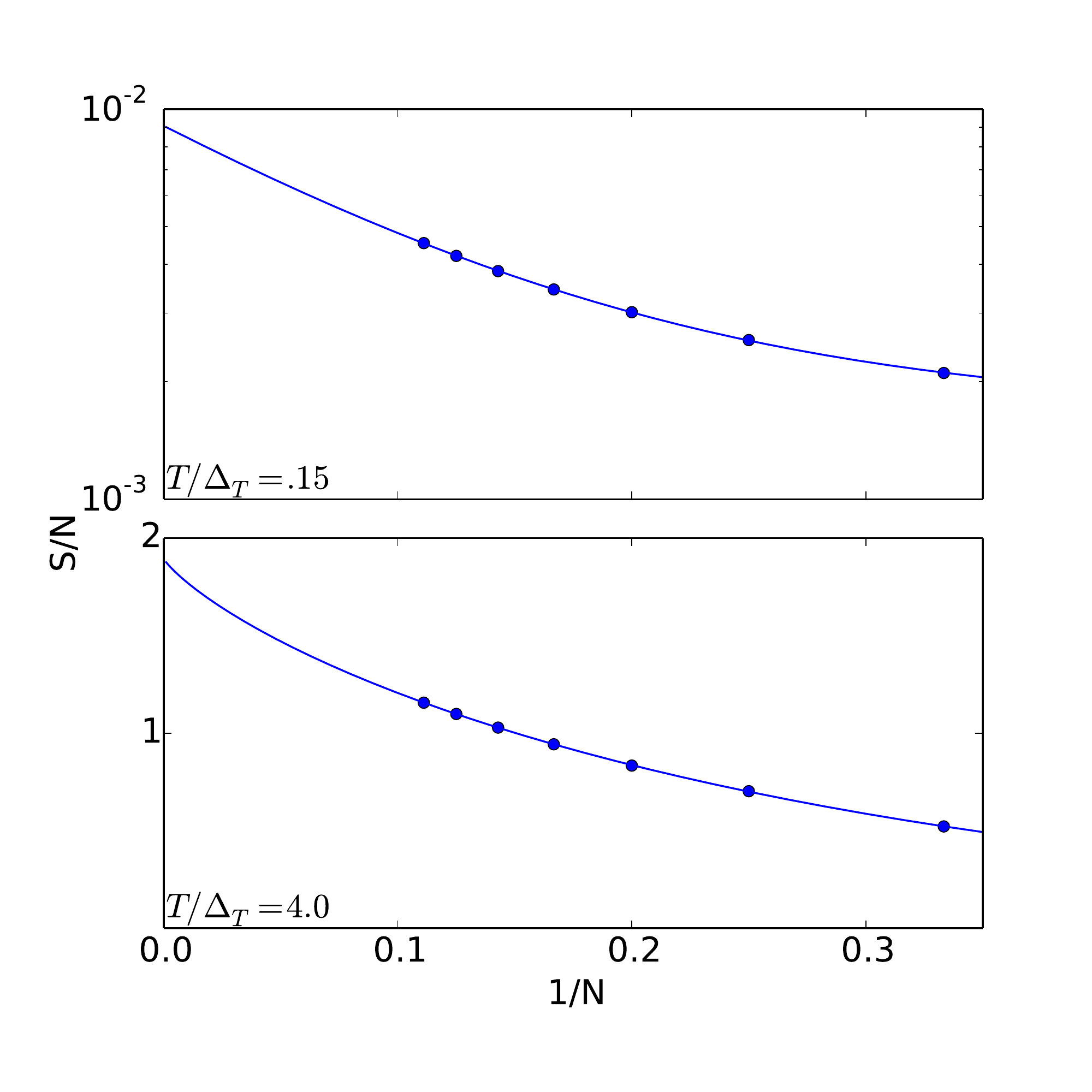}
	\caption{Finite-size extrapolation of the entropy per particle for two different temperatures, $T/\Delta_T=0.15$ (top panel) and  $T/\Delta_T=4$ (bottom panel).  The symbols show results from exact diagonalization for bosons at $\nu=1/2$ and the solid lines are fitting functions, Eq.~\eqref{eq_lowTS} (top panel) and Eq.~\eqref{eq_highTS} (bottom panel).}
	\label{plot_SNextrap}
\end{figure}

This section discusses finite size scaling forms used for the cumulants and the entropy. 
To extrapolate the cumulants to the $N\rightarrow\infty$ limit we use different scaling functions.  The functions are used to fit numerical data.

We use three different scaling functions for the cumulants.  For $\kappa_0$, which is the log of the dimension size, we use Stirling's formula to extract the following large $N$ scaling:
\begin{eqnarray}
\frac{\kappa_0}{N}\sim a+ \frac{b}{N} + \frac{c}{2N}\log{\left(\frac{1}{N}\right)},
\label{eq_s_kappa0}
\end{eqnarray}
where $a,b$ and $c$ are fitting parameters obtained from fitting our numerical data.  For $\kappa_1$ we use a standard finite size scaling function used for energy extrapolations in FQH studies:
\begin{eqnarray}
\frac{\kappa_1}{N} \sim a + \frac{b}{N} + \frac{c}{N^2},
\label{eq_s_kappa1}
\end{eqnarray}
but for higher order cumulants we find that an exponential scaling yields the least error in fitting:
\begin{eqnarray}
\frac{\kappa_{l>1}}{N} \sim a + b\exp[c/N].
\label{eq_s_kappag}
\end{eqnarray}

Figure~\ref{fig_cumulants} shows example numerical extrapolations for boson cumulants using the stochastic trace method.  Fits for bosons for $\kappa_0/N$ and $\kappa_1/N$ yield 1.899(1) and 1.006(5), respectively, close to our expected values, $\approx1.90954$ and $1.0$, thus confirming that our fitting protocol extrapolates well to $N\rightarrow\infty$.  The $N\rightarrow\infty$ limit we find for cumulants is given in Table~I in the main text.  

\begin{figure}[t]
	\centering
	\includegraphics[scale=.45]{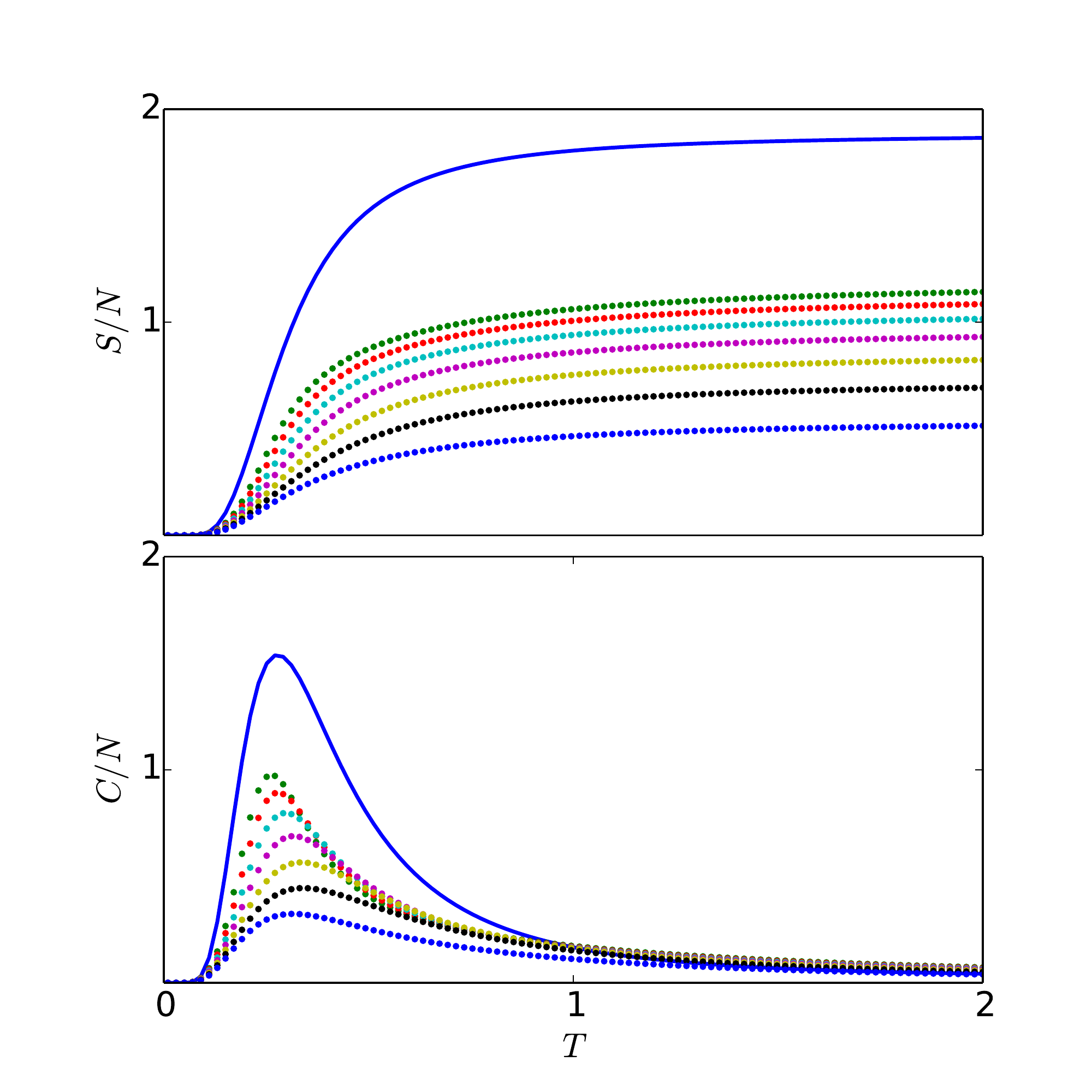}
	\caption{Top panel: Entropy per particle versus temperature from the ansatz, Eq.~(2) in the main text, (solid line) and from exact diagonalization of the Hamiltonian (dotted lines).   The dotted lines ascend with $N$ from $N=3$ to $N=9$. Bottom panel:  The same as the top panel but for the heat capacity per particle. }
	\label{plot_SCAED}
\end{figure}

We also perform finite size scaling of thermodynamic functions directly.  The symbols in the top panel of Fig.~3 in the main text shows the entropy per particle extrapolated to $N\rightarrow\infty$  for several $T$.  To obtain the results presented in Fig.~3 in the main text we performed finite size scaling of the entropy.  The finite size scaling of the entropy per particle is well defined at very low $T$ and very high $T$.  For $T\ll \Delta_T$ we find an exponential scaling:
\begin{eqnarray}
\frac{S}{N}\bigg\vert_{T\ll \Delta_T} \sim \tilde{a}(T) + \tilde{b}(T)\exp[\tilde{c}(T)/N],
\label{eq_lowTS}
\end{eqnarray}
where the tilde indicates temperature dependence in the fitting parameters $\tilde{a}, \tilde{b},$ and $\tilde{c}$.  The high temperature $N$-scaling of the entropy is different.  For $T\gg \Delta_T$ Sterling's formula leads to a log-linear scaling for the entropy:
\begin{eqnarray}
\frac{S}{N}\bigg\vert_{T\gg \Delta_T}  \sim \tilde{a}(T)+  \frac{\tilde{b}(T)}{N} + \frac{\tilde{c}(T)}{2N}\log{\left(\frac{1}{N}\right)}.
\label{eq_highTS}
\end{eqnarray}
Here we note that at infinite $T$ the entropy becomes just the log of the Hilbert space size, $\kappa_0$.  We have checked that our finite size scaling form for the entropy extrapolates to the log of the Hilbert space size for infinite $T$. 

\begin{figure}[t]
	\centering
	\includegraphics[scale=.45]{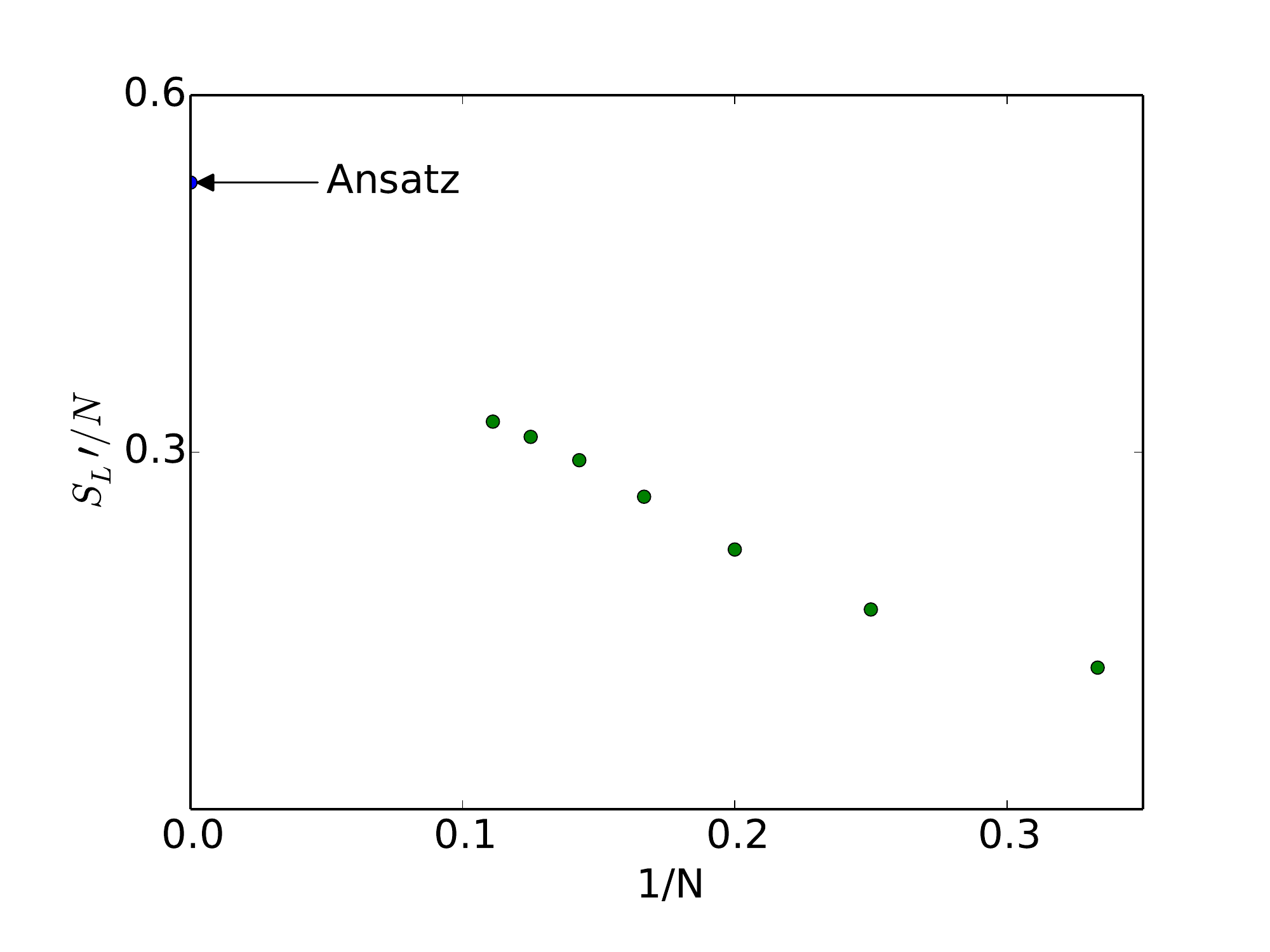}
	\caption{The circles plot the entropy per particle at the inflection points of the entropy versus temperature found in the top panel in Fig.~\ref{plot_SCAED} for each $N$.   The arrow on the $y$-axis plots the entropy per particle at the inflection point obtained from Eq.~(2) in the main text.  Here we see that exact diagonalization results for the inflection point trend towards the ansatz. }
	\label{plot_inflection}
\end{figure}

Fig.~\ref{plot_SNextrap} shows example $N$ scalings for the entropy for two different temperatures.  The top panel uses a log-linear plot to show an extrapolation using the exponential form, Eq.~\eqref{eq_lowTS}, for a characteristic low $T$.  The bottom panel shows an extrapolation using the log-linear form, Eq.~\eqref{eq_highTS}, for a characteristic high $T$.  In both panels we see that the finite size numerical data fall on the extrapolation lines. Intermediate temperatures ($T\sim\Delta_T$) were not accessible because the fitting functions fail to capture numerical data.  We are therefore limited to low and high $T$ results for this method.  In these low and high $T$ regimes we then use the extrapolated values of the entropy to compare with the ansatz, all in the $N\rightarrow\infty$ limit.  Specifically, the symbols in the top panel of Fig.~3 in the main text plot the $y$-intercept obtained from fits at different temperatures as shown in Fig.~\ref{plot_SNextrap}.

\subsection{Intermediate Temperature Trends}

We now turn to intermediate temperatures.  We compare thermodynamic functions obtained from the ansatz (in the thermodynamic limit) and from exact diagonalization (in finite sized systems).  In spite of a lack of a known finite-size scaling form for the thermodynamic functions at intermediate temperatures, we do see that finite size systems do trend to the ansatz constructed in the thermodynamic limit.  

Figure~\ref{plot_SCAED} plots both the entropy and heat capacity over the entire temperature range for both the ansatz and exact diagonalization.  Here we see that the exact diagonalization results trend toward the ansatz.  The location of the peak of the heat capacity shows finite size effects.  But overall the peak of the small system size results approaches  the peak from the ansatz.

To better quantify the trend we consider a single point on the entropy curve.  We define the entropy at which the curvature changes as a function of temperature (the entropy inflection point) to be $S_L^{'}$.  The circles in Fig.~\ref{plot_inflection} plot the entropy at the inflection point obtained from exact diagonalization for several system sizes.  The circles trend toward the value found for $S_L^{'}$ obtained from the ansatz in the thermodynamic limit.

\end{document}